    \documentclass{JHEP3}
    \usepackage{amsmath}
    \usepackage{amssymb}
    \usepackage{graphicx}
    \numberwithin{equation}{section}

    \usepackage{epsfig}

    \newcommand{\Ad}{AdS_4\times {\mathbb C \mathbb P}^3}

    \renewcommand{\[}{\left[}
    \renewcommand{\]}{\right]}
    \renewcommand{\(}{\left(}
    \renewcommand{\)}{\right)}

    \newcommand{\ph}{{\rm ph}}
    \newcommand{\mir}{{\rm mir}}
    \newcommand{\cut}{{\circlearrowright\hspace{-3.4mm}*\hspace{0.2mm}}}
    
    \newcommand{\YF}{Y_{\fF}}
    \newcommand{\Yf}{Y_{\ff}}
    \newcommand{\Ym}[1]{Y_{\fm_#1}}
    
    \newcommand{\Yb}[1]{Y_{\fb_#1}}
    \newcommand{\Yp}[1]{Y_{\fp_#1}}

    \newcommand{\ff}{\scalebox{0.67}{\begin{picture}(5.21,0.0)\put(-1.5,1.2){$\oplus$}\end{picture}}}
    \newcommand{\fF}{\scalebox{0.67}{\begin{picture}(5.21,0.0)\put(-1.5,1.2){$\otimes$}\end{picture}}}
    \newcommand{\fm}{\scalebox{1.15}{\begin{picture}(2.8,0.0)\put(-1.2,-0.33){$\bullet$}\end{picture}}}
    
    \newcommand{\fb}{\scalebox{0.5}{\begin{picture}(7.27,0.3)\put(-1.7,2.65){$\bigcirc$}\end{picture}}}
    \newcommand{\fp}{\scalebox{0.6}[0.7]{\begin{picture}(7.24,0.2)\put(-1.5,0.5){$\bigtriangleup$}\end{picture}}}

    \newcommand{\beq}{\begin{equation}}
    \newcommand{\eeq}{\end{equation}}
    \newcommand\beqa{\begin{eqnarray}}
    \newcommand\eeqa{\end{eqnarray}}
    \newcommand\bea{\begin{array}}
    \newcommand\eea{\end{array}}

    \newcommand\IM{{\rm Im}\,}

    \def\XXint#1#2#3{{\setbox0=\hbox{$#1{#2#3}{\int}$}
    \vcenter{\hbox{$#2#3$}}\kern-.5\wd0}}

    \def\cw{{\hspace{-2.5mm}\unitlength 0.1in
    \begin{picture}(1.00,1.00)(8.67,-11.50)
    \special{pn 8}%
    \special{pa 1120 1100}%
    \special{pa 1100 1120}%
    \special{pa 1080 1100}%
    \special{fp}%
    \end{picture}%
    \hspace{-0mm}}}

    \newcommand{\nn}{\nonumber}

    \newcommand{\COMMENT}[1]{}
    
    \newcommand{\neqa}{\nonumber\end{eqnarray}}
    \newcommand{\la}[1]{\label{#1}}

    \newcommand{\eq}[1]{(\ref{#1})}

    \newcommand{\half}{\frac{1}{2}}

    \renewcommand{\d}{\partial}
    
    \newcommand{\<}{{\langle}}
    \renewcommand{\>}{{\rangle}}

    \newcommand{\re}{\relax{\rm I\kern-.18em R}}

    \newcommand{\rb}{\right)}
    \newcommand{\lb}{\left(}

    \def\su2{{SU(2)}}

    \def\eps{{\epsilon}}

    \def\<{\langle}
    \def\>{\rangle}

    \def\i2{\frac{i}{2}}

    \def\e{{\rm e}}

    \newcommand{\figf}{\scalebox{0.67}{\begin{picture}(5.21,0.0)\put(-1.5,1.2){$\oplus$}\end{picture}}}
    \newcommand{\figF}{\scalebox{0.67}{\begin{picture}(5.21,0.0)\put(-1.5,1.2){$\otimes$}\end{picture}}}
    \newcommand{\figm}{\scalebox{1.15}{\begin{picture}(2.8,0.0)\put(-1.2,-0.33){$\bullet$}\end{picture}}}
    
    \newcommand{\figb}{\scalebox{0.5}{\begin{picture}(7.27,0.3)\put(-1.7,2.65){$\bigcirc$}\end{picture}}}
    \newcommand{\figp}{\scalebox{0.6}[0.7]{\begin{picture}(7.24,0.2)\put(-1.5,0.5){$\bigtriangleup$}\end{picture}}}

    \newcommand{\fo}{\scalebox{0.67}{\begin{picture}(5.21,0.0)\put(-1.5,1.2){$\blacktriangleleft$}\end{picture}}}
    \newcommand{\fO}{\scalebox{0.67}{\begin{picture}(5.21,0.0)\put(-1.5,1.2){$\blacktriangleright$}\end{picture}}}

    \newcommand{\be}{\begin{equation}}
    \newcommand{\ee}{\end{equation}}
    \newcommand{\bg}{\begin{gather}}
    \newcommand{\eg}{\end{gather}}
    \newcommand{\bseq}{\begin{subequations}}
    \newcommand{\eseq}{\end{subequations}}
    \def\half{\frac{1}{2}}

    \newcommand{\bac}{\begin{array}{l}}
    \newcommand{\eac}{\end{array}}

    \newcommand{\bal}{\begin{array}{l}}
    \newcommand{\eal}{\end{array}}

    \newcommand{\cQ}{\mathcal{Q}}

    \title{Y-system, TBA and Quasi-Classical Strings  in $\Ad$}
    \author{
	Nikolay Gromov\\ Mathematics Department, King's College London,\\
      The Strand, London WC2R 2LS, UK \& \\
   DESY Theory, Notkestr. 85 22603 Hamburg, Germany \& \\
   II. Institut f\"{u}r Theoretische Physik Universit\"{a}t Hamburg,
      Luruper Chaussee 149 22761 Hamburg Germany \& \\
   St.Petersburg INP, Gatchina, 188 300, St.Petersburg, Russia\\
    E-mail: \email{nikgromov$\bullet$gmail.com}}
    \author{Fedor~Levkovich-Maslyuk\\ Physics Department, Moscow State University, 119991, Moscow, Russia\\
    E-mail: \email{fedor.levkovich$\bullet$gmail.com}}

    \abstract{
        We study the exact spectrum of the ${\rm AdS}_4/{\rm CFT}_3$
    duality put forward by Aharony, Bergman, Jafferis and Maldacena (ABJM).
    We derive thermodynamic Bethe ansatz (TBA) equations for the planar ABJM
    theory, starting from ``mirror" asymptotic Bethe equations which we conjecture.
    We also propose generalization of the TBA equations for excited states. The recently proposed Y-system is completely consistent with the TBA
    equations
    for a large subsector of the theory, but should be modified in general.
    We find the general asymptotic infinite length solution of the Y-system,
    and also several solutions to all wrapping orders in the strong coupling scaling limit.
    To make a comparison with results obtained from string theory, we assume that the all-loop Bethe ansatz of N.G. and P. Vieira
    is the valid worldsheet theory description in the asymptotic regime.
    In this case we find complete agreement, to all orders in wrappings, between the
    solution of our Y-system and generic quasi-classical string spectrum in ${\rm AdS}_3\times {\rm S}^1$.
    }

    \preprint{}
    
\begin{document}


    \section{Introduction}
    The AdS/CFT correspondence \cite{Maldacena:1997re} continues to
    be a source of exciting new results in gauge
    and string theories. The best-studied example of the duality
    is the correspondence between four-dimensional ${\cal N}=4$
    super Yang-Mills (SYM) theory and Type IIB superstring theory on
    ${\rm AdS}_5\times {\rm S}^5$. Another example is the recently found duality between
    Type IIA string theory on ${\rm AdS}_4\times {\mathbb{CP}}^3$
    and three-dimensional ${\cal N}=6$ super Chern-Simons (SCS) theory \cite{Aharony:2008ug}.

    Remarkably, evidence for integrability has been found both in
    the gauge theory \cite{Minahan:2002ve,MSS} and in the string theory \cite{Bena:2003wd,AF2} in the planar limit
    of large number of colors.
    In SYM
    further intensive development \cite{St,KMMZ} has led to complete description of anomalous dimensions
    of infinitely long operators by means of the Asymptotic Bethe Ansatz (ABA) equations
    \cite{Beisert:2005fw,Beisert:2005fw2}.
    Similar equations were found in \cite{GV2,AN1} for SCS.
    Very recently the integrability approach was also extended to ${\rm AdS}_3/{\rm CFT}_2$ dual pairs
    \cite{Babichenko:2009dk}.

    For complete solution of the planar AdS/CFT spectral problem
    one should be able to solve the integrable two dimensional worldsheet
    theory in finite volume. The program of applying the methods of relativistic integrable
    field theories for finite size spectrum of AdS/CFT was started in \cite{Ambjorn:2005wa}.
In \cite{Bajnok:2008bm} a generalization of the
    L\"uscher type formula was proposed for the first finite volume
    correction to the asymptotic spectrum generated by ABA.
    This information, as well as experience with relativistic
    integrable theories \cite{Zamolodchikov:1991et},
    led to the Y-system proposed in \cite{Gromov:2009zz} for exact solution
    of both ${\rm AdS}_5\times {\rm S}^5$ and ${\rm AdS}_4\times {\mathbb{CP}^3}$ theories.
    As we show in this work, the proposal of \cite{Gromov:2009zz} for ABJM theory is only valid
    in a certain large subsector of the theory,
    and should be modified to describe the general case.    \FIGURE[ht]{\la{Fig:Ys}
    \begin{tabular}{cc}
    \includegraphics[scale=0.4]{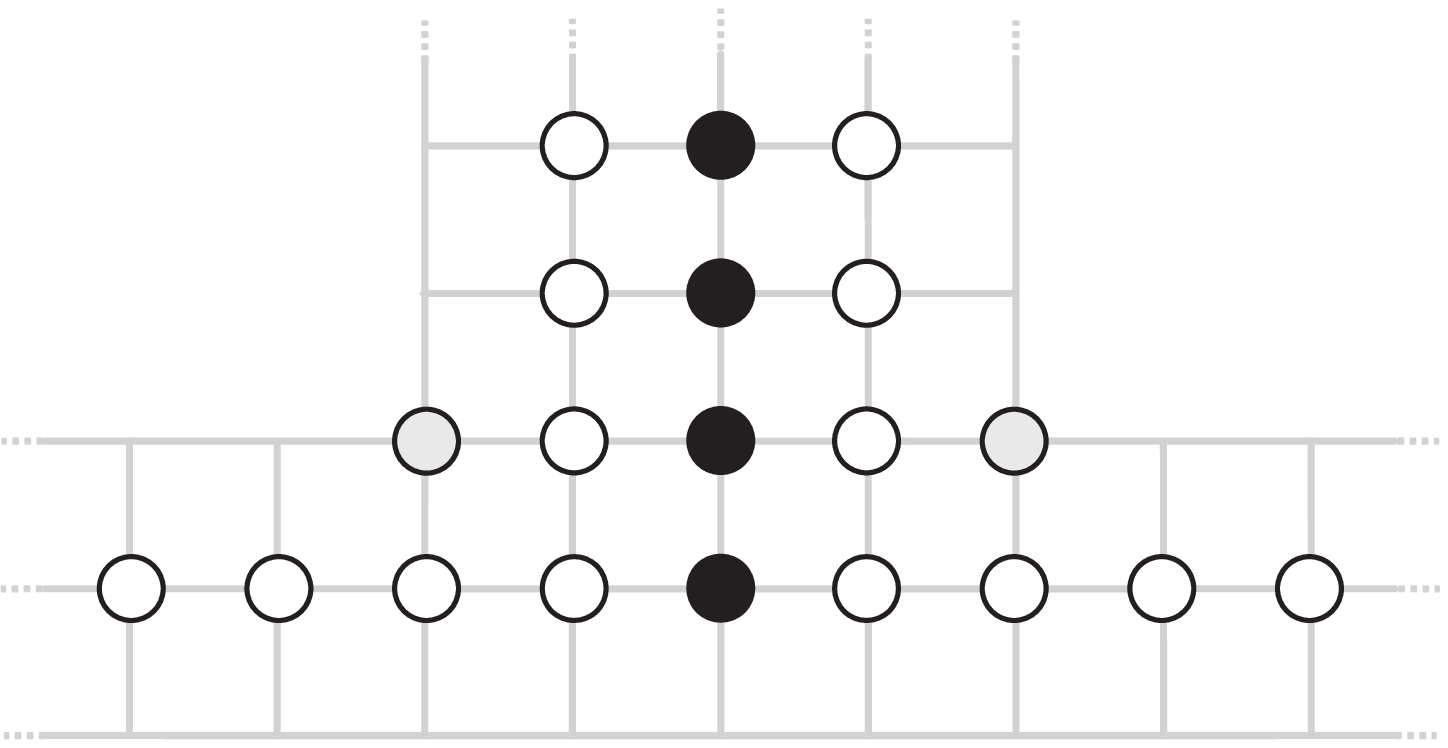}
    &
    \includegraphics[scale=0.4]{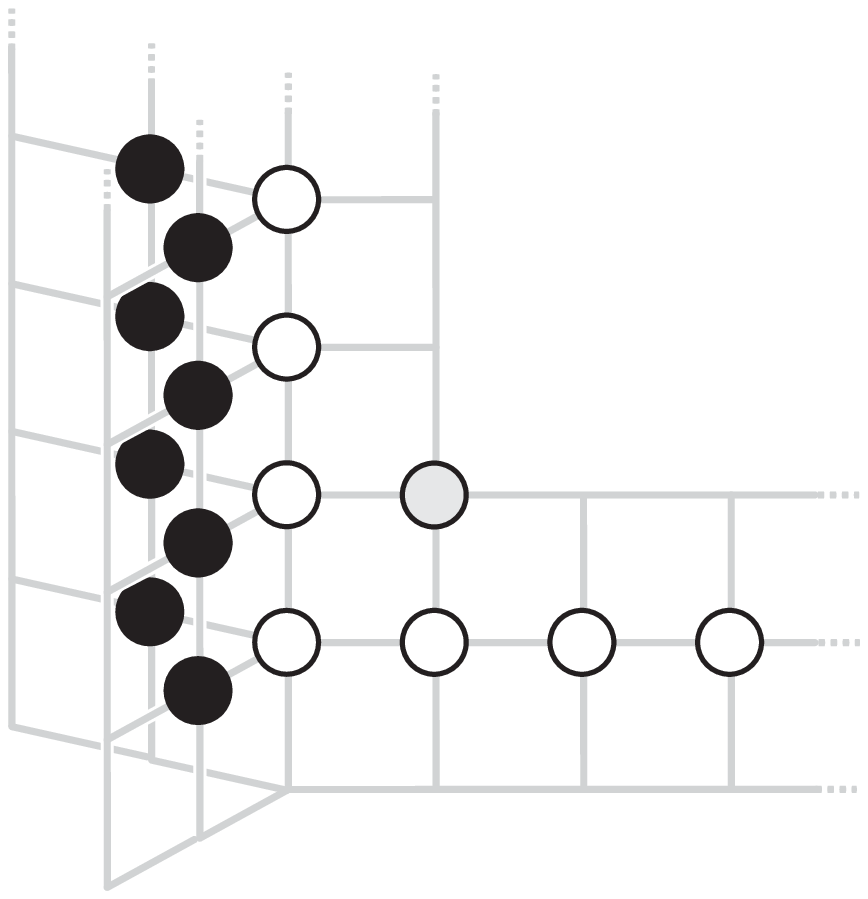}
    \\
    ${{\rm AdS}_5\times {\rm S}^5}$
    &
    ${{\rm AdS}_4\times {\mathbb{CP}}^3}$
    \end{tabular}
    \caption{Graphical representation of the Y-systems \cite{Gromov:2009zz}.
    Circles correspond to Y-functions.
    Black nodes are the ``massive" nodes which are suppressed
    for asymptotically large length $L$.
    For gray circles in the corners the equation
    cannot be written ``locally" in terms of Y's.}
    \label{AdS4Ysys}
    }

    A graphical representation of the Y-systems of \cite{Gromov:2009zz} is given in Fig.~\ref{Fig:Ys}
    where the Y-functions are represented by circles.
    Each value of the index $A$, which labels the Y-functions (functions of the spectral parameter $u$), corresponds to a
    node of this diagram. For each node $A$, except the gray ones, the Y-system equation has the form
    \be\la{genY}
        Y_A^+ Y_A^- = \frac{\prod_B (1 + Y_B)}{\prod_C (1 + 1/Y_C)},
    \ee
    where $Y_A^\pm=Y_A(u\pm i/2)$ and the index $B$ (resp. $C$) labels the nodes
    connected to the $A$ node by horizontal (resp. vertical) lines.\footnote{For the gray node the equations cannot be written
    as functional equations in terms of $Y$'s. In many cases it is convenient to
    parameterize the Y-functions in terms of T-functions, which satisfy the Hirota functional equation.
    The ``non-local" equation for the gray nodes is replaced (see for example \cite{Gromov:2009zz}) by a ``local" one in terms of T-functions.}

    In this paper we argue that in the AdS$_4$/CFT$_3$ case equation \eq{genY} for black nodes (Fig.~\ref{Fig:Ys}, on the right) should be replaced
    by rather unusual equations
    \beqa
    \label{YoOeq1}
        Y_{{\fO}_{a}}^+Y_{{\fo}_{a}}^-
        &=& \frac {1+Y_{\figp_a}}{(1+1/Y_{\fo_{a+1}})(1+1/Y_{\fO_{a-1}})} \,\, , \,\, a>1,\\
     \label{YoOeq2}
     {Y_{{\fo}_{a}}^+Y_{{\fO}_{a}}^-}
        &=& \frac {1+Y_{\figp_a}}{(1+1/Y_{\fO_{a+1}})(1+1/Y_{\fo_{a-1}})} \,\, , \,\, a>1,
    \eeqa
    \beq
    \label{YoOeq3}
        Y_{{\fO}_{1}}^+Y_{{\fo}_{1}}^-
        = \frac {1+Y_{\fF}}{(1+1/Y_{\fo_{2}})} \,\, ,\;\;
         {Y_{{\fo}_{1}}^+Y_{{\fO}_{1}}^-}
        = \frac {1+Y_{\fF}}{(1+1/Y_{\fO_{2}})} \,\, ,
   \eeq
while all the other Y-system equations of \cite{Gromov:2009zz} need not be changed.\footnote{
    With the following identification between Y-functions of \cite{Gromov:2009zz} and new Y-functions: $Y^{4}_{a,0} = Y_{{\fo}_{a}}, Y^{\bar 4}_{a,0} = Y_{{\fO}_{a}}$}
    Notice that for the case $Y_{{\fO}_{a}}=Y_{{\fo}_{a}}$ the new equations \eq{YoOeq1}--\eq{YoOeq3} coincide with the ones originally proposed in \cite{Gromov:2009zz}.

    Once the Y-functions are found the energy of the state can be computed from
    \beq
    \label{Eintro}
    E=\sum\int_{-\infty}^{\infty}\frac{du}{2\pi i}\frac{\d \epsilon_a^{\rm mir}(u)}{\d u}
    \log(1+Y^{\mir}_{\fO_a})(1+Y^{\mir}_{\fo_a})
    +\sum_{j=1}^{K_4}\epsilon^{\rm ph}(u_{4,j})
        +\sum_{j=1}^{K_{\bar 4}}\epsilon^{\rm ph}(u_{\bar 4,j})\;,
    \eeq
    where $u_j$ are the exact Bethe roots given by
    \beq
    \label{Yismin1}
    Y^{\ph}_{\fo_1}(u_{4,j})=-1\;\;,\;\;
        Y^{\ph}_{\fO_1}(u_{\bar 4,j})=-1\;,
    \eeq
    and $\epsilon$ is the single magnon dispersion introduced in \eq{peps} (see section 3.3 for more details).

    In the ${\rm AdS}_5$ case
    the Y-system passes some nontrivial tests --
    in \cite{Gromov:2009zz} the $4$-loop perturbative
    result \cite{Fiamberti:2008sh} was reproduced\footnote{Technically the derivation of  \cite{Gromov:2009zz} is very similar to \cite{Janik4loops}, where the 4-loop
    perturbative results were reproduced for the first time.},
    and more recently a comparison was made at
    $5$ loops in \cite{Fiamberti:2009jw}. In
    \cite{Bombardelli:2009ns,Gromov:2009bc,Arutyunov:2009ur}
    the Y-system was also shown to be consistent with the thermodynamic Bethe ansatz (TBA) approach\footnote{
    In \cite{Arutyunov:2009ur} the Y-system was only obtained in the interval $-2g<u<2g$.
    At the same time the authors of \cite{Arutyunov:2009ur} failed to get the
    Y-system of \cite{Gromov:2009zz} for $|u|>2g$ and the discrepancy was stated.
    The reason of this misunderstanding is that some of the Y-functions
    have branch points at $u=\pm 2g\pm \tfrac{i}{2}$ with branch cuts going to $\pm \infty\pm\tfrac{i}{2}$, parallel to the real axis. Thus for real $u$ the quantity
    $Y(u+\tfrac{i}{2})Y(u-\tfrac{i}{2})$ can be understood for instance as $Y(u+\tfrac{i}{2}-i0)Y(u-\tfrac{i}{2}+i0)$ or
    $Y(u+\tfrac{i}{2}+i0)Y(u-\tfrac{i}{2}+i0)$. The first prescription (chosen in \cite{Arutyunov:2009ur}) leads to the discrepancy
    whereas the second does not. Note that the problem is only present for real $u$ i.e. for measure zero subset of the complex plane.
    Any prescription which preserves continuity leads to agreement with \cite{Gromov:2009zz}. In
    \cite{Frolov:2009in} (after private communication with P.Vieira) the issue was resolved.}
    .

    The TBA equations, describing the ground state energy, do not lead to
    any nontrivial dependence of that energy on the coupling since the ground state is protected
    by super-symmetries. In \cite{Gromov:2009bc} an extension of these equations
    was proposed to describe the excited states. These equations were solved
    numerically in \cite{Gromov:2009zb} for the first non-trivial Konishi operator
    \cite{Bianchi:2001cm},
    giving for the first time the anomalous dimension of a non-protected operator in a wide range of values of the 't Hooft coupling $\lambda$
    for a 4D gauge theory in the planar limit.
    The numerical results also indicate agreement with the string prediction \cite{GKP}\footnote{
    Much later the equations of \cite{Gromov:2009zb} were rederived by another group \cite{Arutyunov:2009ax}.
    The authors of \cite{Arutyunov:2009ax} confirmed the validity of the equations at least
    in the range of the coupling $0<\lambda<700$ where a numerical solution was obtained.
    The perturbation theory for the world-sheet sigma model in the formulations of \cite{Berkovits:2000fe}
    is naturally organized in powers of $1/\sqrt{\lambda}$ and thus the value $\lambda\sim 700$
    should already give the asymptotic of $E(\lambda)$ with a good precision especially when
    an appropriate extrapolation procedure is applied. This holds assuming the analyticity of $E(\lambda)$ for real positive values of $\lambda$
    which is however doubted in \cite{Arutyunov:2009ax}.
     }. The results of  \cite{Gromov:2009zb} disagree in the sub-leading $1/\lambda^{1/4}$ order with two
     string computations \cite{Roiban:2009aa} and \cite{Arutyunov:2005hd}
     which also disagree with each other and are based on rather strong assumptions.
     In \cite{Arutyunov:2005hd} a truncated model is considered whereas
     \cite{Roiban:2009aa} assumes the applicability of the quasiclassics    in the small charge limit.
     We hope that a first principles calculation can be done using Berkoviz's pure spinor
     formalism \cite{Berkovits:2000fe}.

    Another very recent test of the Y-system of \cite{Gromov:2009zz} for ${\rm AdS}_5/{\rm CFT}_4$
    was done at strong coupling \cite{Gromov:2009tq}. An {\it analytical} solution of the Y-system was found
    for generic classical string motion inside ${\rm AdS}_3\times {\rm S}$.
    It was shown to agree with the quasi-classical one-loop spectrum to all orders in wrapping
    providing thus a deep structural test of the Y-system in the regime where the
    ABA fails completely.

    In this paper we apply the technique of TBA
    for the ${\rm AdS}_4\times{\mathbb{CP}^3}$ theory
    to test the Y-system we propose.
    We also present the general
    asymptotic infinite length solution of the Y-system.
    The asymptotic solution is very important since it allows to establish a correspondence between the exact
    solution of the Y-system and the physical states of the theory. It can be also used at weak coupling where it is a good approximation to study the leading wrapping effects.
    In addition, we find strong
    coupling solutions of the Y-system in two cases and compare results with the
    quasi-classical string spectrum thus testing deeply the structure of the Y-system to all orders in wrapping.


    \section{Asymptotic large $L$ solution of Y-system}
    The asymptotic spectrum of the theory can be found using
    asymptotic Bethe ansatz (ABA) techniques. In this section we describe the
    ABA equations of \cite{GV2} and link them with the Y-system formalism
    by presenting the general asymptotic solution of the Y-system. That solution extends the one of \cite{Gromov:2009zz}.

    In the asymptotic regime the counting of the states is very clear and well established.
    One can analytically continue the solution of the Y-system from the asymptotic
    regime, where the solution is explicit, to finite volume. Usually this continuations is unique
    (see for example \cite{Gromov:2008gj}) and allows to fix the solution of Y-system.
    Technically at the moment it is not known how to perform this procedure for the general excited state
    in AdS/CFT. We show how to apply this general method
    \cite{Gromov:2008gj} for the ``${\mathfrak{sl}(2)}$" subsector and also at strong coupling.

    \subsection{Asymptotic Bethe ansatz equations for physical ${\rm AdS}_4/{\rm CFT}_3$}

    Here we present the asymptotic Bethe equations
    for the ${\rm AdS}_4/{\rm CFT}_3$ theory,
    which were for the first time obtained in~\cite{GV2}.
    We will also introduce some notation useful for the sequel.

    First we define the Zhukowski variable $x(u)$:
    \be
    \label{xdef}
        x + \frac{1}{x} = \frac{u}{h(\lambda)},
    \ee
    where $h(\lambda)$ is some unknown function of the 't Hooft coupling $\lambda$.
    It should have the following asymptotics at weak coupling and strong coupling:
    \beq
    h(\lambda)=\lambda+h_3\lambda^3+{\cal O}\(\lambda^5\)=
    \sqrt{\lambda/2}+h^0+{\cal O}\(\frac{1}{\sqrt{\lambda}}\)\;.
    \eeq
    Recently the coefficient $h_3=-8+2\zeta_2$ was computed directly from the Super--Chern--Simons
    perturbation theory \cite{MSS}. At strong coupling
    the situation is less clear: in \cite{Sh} and \cite{Bergman:2009zh}
    the coefficient $h^0$ was argued to be $0$
    whereas in \cite{McLoughlin:2008he} some evidence was given
    in favor of a different value $-\frac{\log2}{2\pi}$ (see also \cite{Bombardelli:2008qd}).
    Hopefully this issue could be analyzed from world-sheet sigma model first principles calculation
    like in \cite{Mazzucato:2009fv}.

    Equation~\eq{xdef} admits two solutions, and we define two branches of the function $x(u)$, which are called ``mirror" and ``physical":
    \beq
    \label{xBranches}
        x^{\ph}(u)=\frac{1}{2}\lb
        \frac{u}{h}+\sqrt{\frac{u}{h}-2}\;\sqrt{\frac{u}{h}+2}\rb\;\;,\;\;
        x^{\mir}(u)=\frac{1}{2}\lb \frac{u}{h}+i\sqrt{4-\frac{u^2}{h^2}}\rb \,.
    \eeq
    Here, by $\sqrt{u}$ we denote the principal branch of the square root. This definition of mirror and physical branches
    is the same as in the ${\rm AdS}_5/{\rm CFT}_4$ case~\cite{Arutyunov:2007tc, Gromov:2009bc}, with
    the ${\rm AdS}_5/{\rm CFT}_4$ coupling $g$ replaced by $h(\lambda)$. Above the real axis, the
    mirror and physical branches coincide. $x^{\ph} (u)$ is obtained by analytical
    continuation from the upper half plane to the plane with the cut $(-2h,2h)$,
    and $x^{\mir} (u)$ -- by continuation to the plane with the cut $(-\infty, -2h)\cup (2h,+\infty)$.
    The ${\rm AdS}_4/{\rm CFT}_3$ Bethe equations~\cite{GV2} for the original (physical) theory are written in
    terms of $x^\ph (u)$, while the mirror
    Bethe equations we conjecture include $x^\mir (u)$,
    in analogy with the ${\rm AdS}_5/{\rm CFT}_4$ case \cite{Arutyunov:2007tc} (see section 3). In sections 4 and 5 we use the mirror branch of $x$ if its argument is a free variable, and the physical branch for $x(u_j)$, with $u_j$ being the Bethe roots.

    In the physical ABA equations of~\cite{GV2} there are five types of Bethe roots: $u_1,u_2,u_3,u_4$ and $u_{\bar 4}$. Conserved local
    charges (the heights Hamiltonians) in ${\rm AdS}_4/{\rm CFT}_3$ are expressed in terms of the momentum-carrying roots $u_4$ and $u_{\bar4}$:
    \be
    {\cal Q}_n= \sum_{j=1}^{K_4} \textbf{q}_n(u_{4,j})+ \sum_{j=1}^{K_{\bar 4}} \textbf{q}_n(u_{\bar 4,j}) \,\,\, , \,\,\, \textbf{q}_n=\frac{i}{n-1} \(\frac{1}{(x^+)^{n-1}}-\frac{1}{(x^-)^{n-1}}\),
    \ee
    where we have used general notation
    \be
        f^\pm(u) \equiv f(u\pm i/2), \ \ \ f^{[+a]} \equiv f(u + ia/2).
    \ee
   In particular, string state energies in $\Ad$ or operator anomalous dimensions in the dual gauge theory are obtained from
    $E = h(\lambda){\cal Q}_2$.

    The momentum and energy which correspond to a single Bethe root $u_4$ or $u_{\bar4}$ are given by
    \be \la{peps}
        p = \frac{1}{i}\log\frac{x^+}{x^-},
        \ \ \ \epsilon = \half + h(\lambda) \( \frac{i}{x^+} - \frac{i}{x^-}\),
    \ee
    and the charge $\cQ_1$ is the sum of all momenta:
    \be
        \mathcal{Q}_1 = \sum_{j=1}^{K_4} p(u_{4,j})+ \sum_{j=1}^{K_{\bar 4}} p(u_{\bar 4,j}).
    \ee
    To write the Bethe equations in compact form, we introduce the following notation:
    \beq
    \label{defR}
    R_l^{(\pm)}=\prod_{j=1}^{K_l}\frac{x(u)-x_{l,j}^{\mp}}{(x_{l,j}^\mp)^{1/2}}\;\;,\;\;
    R_l=\prod_{j=1}^{K_l}({x(u)-x_{l,j}})\;\;,\;\;
    \eeq
    \beq
    \label{defB}
    B_l^{(\pm)}=\prod_{j=1}^{K_l}\frac{1/x(u)-x_{l,j}^{\mp}}{(x_{l,j}^\mp)^{1/2}}\;\;,\;\;
    B_l=\prod_{j=1}^{K_l}(1/x(u)-x_{l,j})\;\;,\;\;
    \eeq
    \be
    \label{defQS}
        Q_l(u) \equiv \prod_{j=1}^{K_l}(u-u_{l,j}),
        \ \ \
        S_l(u) \equiv \prod_{j=1}^{K_l} \sigma_{\rm BES} (x(u), x_{l,j})
    \ee
    where $\sigma_{\rm BES}$ is the Beisert-Eden-Staudacher dressing kernel~\cite{GV2}.
    The Bethe equations of~\cite{GV2} in $\mathfrak{sl}_2$ favored grading have the form\footnote{Here, as well as when constructing the asymptotic solution of Y-system, one should be careful with the sign ambiguity in the square root factors inside $\e^{\half i\cQ_1}$ and $B_l, R_l$.}
    \beqa
    \nn
        +1 &=&
        \e^{-\half i \cQ_1}
        \left. \frac{Q_2^+ B_{4}^{(-)} B_{\bar 4}^{(-)}}
                    {Q_2^- B_{4}^{(+)} B_{\bar 4}^{(+)}} \right|_{u_{1,k}}
        \;,\\ \nn
        -1 &=&
        \left. \frac{Q_2^{--} Q_1^+ Q_3^+}
                    {Q_2^{++} Q_1^- Q_3^-} \right|_{u_{2,k}}\;,\\
        +1 &=&
        \e^{\half i \cQ_1}
        \left. \frac{Q_2^{+} R_{4}^{(-)} R_{\bar 4}^{(-)}}
                    {Q_2^{-} R_{4}^{(+)} R_{\bar 4}^{(+)}} \right|_{u_{3,k}}\;,\la{ABA}
    \\  \nn
        +1 &=&
        \e^{\half i \cQ_1} \e^{-Lip(u_{4,k})}
        \left. \frac{B_{1}^{+} R_{3}^{+}Q_4^{++} R_{4}^{-(-)} R_{\bar4}^{-(-)}}
                    {B_{1}^{-} R_{3}^{-}Q_4^{--} R_{4}^{+(+)} R_{\bar4}^{+(+)}}
        S_4 S_{\bar 4}\right|_{u_{4,k}}\;,
    \\  \nn
        +1 &=&
        \e^{\half i \cQ_1} \e^{-Lip(u_{\bar4,k})}
        \left. \frac{B_{1}^{+} R_{3}^{+}Q_{\bar4}^{++} R_{4}^{-(-)} R_{\bar4}^{-(-)}}
                    {B_{1}^{-} R_{3}^{-}Q_{\bar4}^{--} R_{4}^{+(+)} R_{\bar4}^{+(+)}}
        S_4 S_{\bar 4}\right|_{u_{\bar4,k}}\;,
    \eeqa
    where $L$ is the length of the effective spin chain, and corresponds to the string momentum or length of the operator in the CS theory. The above equations describe the spectrum correctly in the limit $L\to\infty$. We stress again that in those equations the physical branch $x^{\ph}$ of the function $x$ should be used in all places, e.g. inside expressions ~\eq{defR},~\eq{defB},~\eq{defQS} for $B_l$, $R_l$ and $S_l$.

    The Bethe roots are additionally constrained by the zero momentum condition
    \beqa\la{zemom}
        1= \prod_{j=1}^{K_4} \frac{x^+_{4,j}}{x^-_{4,j}}
        \prod_{j=1}^{K_{\bar 4}} \frac{x^+_{\bar 4,j}}{x^-_{\bar 4,j}}
        \Leftrightarrow
        \cQ_1 = 2\pi m.
    \eeqa

    \subsection{General asymptotic solution}
    As we mentioned in the beginning of this section the asymptotic (large $L$) solution of the Y-system
    plays an important role in the whole Y-system
    construction. It allows to link a particular solution of the
    Y-system with an actual state of the theory.
    The asymptotic solutions are in one-to-one correspondence with the solutions
    of ABA equations.

    In many cases one can analytically
    continue a solution from asymptotically large volume
    to finite volume.
    In \cite{Gromov:2009tq} another way
    to inject information about the state of the theory was proposed:
    demanding that the exact functions $Y_{as}$ approach
    the formal asymptotic solution for infinite $a$ or $s$\footnote{
    This should give the same result as analytical continuation in $L$.
    Usually, variations of Y's in $L$ vanish at large $a$ and $s$.
    The values of the Bethe roots inside the asymptotic solution should be equal to their exact values
    e.g. $Y^{\ph}_{\fo_1}(u_{4,j})=-1$.
    One should study this point in more detail.}.
    Then one can still use the same counting of the
    states as in the ABA even for finite volumes\footnote{One cannot
    exclude completely that this procedure fails for some particular small volumes.}.
    This prescription was shown to work especially successfully
    in the strong coupling scaling limit \cite{Gromov:2009tq}, which we describe below.

    In view of its importance
    we will review the construction of \cite{Gromov:2009zz} for
    the asymptotic large $L$ solution of the Y-system in this section and extend it to the case
    $Y_{\fo}\neq Y_{\fO}$.
    To distinguish the asymptotic Y functions from the exact ones we
    use the bold font:
    \beq\la{YbtoT}
    {\bf Y}_{\fp_a}=\frac{{\bf T}^+_{a,1}{\bf T}^-_{a,1}}{{\bf T}_{a+1,1}{\bf T}_{a-1,1}}-1\;\;,\;\;
    1/{\bf Y}_{\fb_s}=\frac{{\bf T}^+_{1,s}{\bf T}^-_{1,s}}{{\bf T}_{1,s+1}{\bf T}_{1,s-1}}-1
    \eeq
    \beqa
    &&{\bf Y}_{\fo_a}\simeq \(\frac{x^{[-a]}}{x^{[+a]}}\)^{L}{\bf T}_{a,1}\prod_{n=-\frac{a-1}{2}}^{\frac{a-1}{2}}
    \Phi_4^{\theta^{\rm E}_{na}}(u+in)\Phi_{\bar 4}^{\theta^{\rm O}_{na}}(u+in)\;\;,\;\;
\\
    &&{\bf Y}_{\fO_a}\simeq \(\frac{x^{[-a]}}{x^{[+a]}}\)^{L}{\bf T}_{a,1}
    \prod_{n=-\frac{a-1}{2}}^{\frac{a-1}{2}}
    \Phi_4^{\theta^{\rm O}_{na}}(u+in)\Phi_{\bar 4}^{\theta^{\rm E}_{na}}(u+in)
    \eeqa
    where $\theta^{\rm E}_{na}$ is $0$ for even and $1$ for odd terms in the product:
    \beq\la{eqtheta}
        \theta^{\rm E}_{na}\equiv\left\{
        \begin{array}{ll}
        1,&n+\frac{a-1}{2}{\rm \ is \ even}\\
        0,&n+\frac{a-1}{2}{\rm \ is \ odd}
        \end{array}\right.
    \eeq
    and $\theta^{\rm O}_{na} \equiv 1-\theta^{\rm E}_{na}$. The factors $\Phi_{4}(u)$ and $\Phi_{\bar 4}(u)$ are constructed in such a way that
    the ABA equations \eq{ABA} for the momentum carrying nodes are given by
    ${\bf Y}_{\fo_1}^{\ph}(u_{4,j})=-1$ and ${\bf Y}_{\fO_1}^{\ph}(u_{\bar 4,j})=-1$. This leads to (using that ${\bf T}_{1,1}(u_{4,j})=-Q_3^+/Q_3^-$)
    \beq\la{Phi4}
    \Phi_{4}(u)=S_4 S_{\bar 4}\frac{B_4^{(+)+}R_{\bar 4}^{(-)-}B_1^+B_3^-}{B_4^{(-)-}R_{\bar 4}^{(+)+}B_1^-B_3^+}
    e^{-i{\cal Q}_1/2}\;\;,\;\;
    \Phi_{\bar 4}(u)=S_4 S_{\bar 4}\frac{B_{\bar 4}^{(+)+}R_{4}^{(-)-}B_1^+B_3^-}{B_{\bar 4}^{(-)-}R_{4}^{(+)+}B_1^-B_3^+}
    e^{+i{\cal Q}_1/2}\;.
    \eeq
    The ${\bf T}_{a,s}$ functions which enter the definitions
    of ${\bf Y}_{a,s}$ can be computed from the generating functional \cite{Tsuboi:1997iq,Kazakov:2007fy}
    \beq\la{eqW}
    {\cal W}=\[1-\frac{Q_1^- B^{(+)+}R^{(+)-}}{Q_1^+ B^{(-)+}R^{(-)-}}D\]
    \!\frac{1}{\[1-\frac{Q_3^{+}Q_2^{--}  R^{(+)-}}{\e^{\half i \cQ_1}Q_3^-Q_2  R^{(-)-}}D\]\[1-\frac{Q_1^- Q_2^{++} R^{(+)-}}
    {\e^{\half i \cQ_1}Q_1^+ Q_2 R^{(-)-}}D\]}
    \!\[1-\frac{Q_3^+}{Q_3^-}D\]
    \eeq
    where $D=e^{-i\d_u}$ is the shift operator and $R=R_{4}R_{\bar 4},\;B=B_{4}B_{\bar 4}$. Expansion of this generating functional yields
    eigenvalues of the ${\mathfrak{su}}(2|2)$
    transfer matrices:
    \beq
    {\cal W}=\sum_{s=0}^\infty {\bf T}_{1,s}(u+i\tfrac{1-s}{2})D^s\;\;,\;\;
    {\cal W}^{-1}=\sum_{a=0}^\infty (-1)^a {\bf T}_{a,1}(u+i\tfrac{1-a}{2})D^a\;.
    \eeq

	In Appendix B we also present the expressions for the asymptotic solution after the duality transformation, which exchanges the $sl(2)$ and $su(2)$ sectors. One can see from those formulas that for $u_{4,j}=u_{\bar4,j}$ the asymptotic solution exactly conicides with the one proposed in \cite{Gromov:2009zz}.

    In the next subsection we expand the asymptotic solution in the scaling
    strong coupling limit.

    \subsection{Asymptotic solution in scaling limit}
    The scaling limit is the strong coupling limit
    $\lambda\to\infty$ where the number of Bethe roots $M$ and the operator length $L$
    go to infinity as $\sqrt\lambda$. The Bethe roots $x_j$
    are distributed along cuts ${\cal C}$ on the complex plane $x$ in this limit \cite{Beisert:2003xu}.
    These cuts can be understood as branch cuts of a $10$-sheet Riemann surface which corresponds
    to a certain function.
    One can interpret them as the eigenvalues of the
    classical monodromy matrix, which are usually written as
    $\lambda_a=e^{-i q_a}$, with $q_a$ being the so called quasi-momenta. Similarly to \cite{GV2}
    for the $\eta=-1$ grading we get
    \beqa
    \bea{lll}
    q_2 =& \frac{Lx/h+{\cal Q}_2x}{x^2-1}&+H_1-{\bar H}_4-{\bar H}_{\bar 4}+{\bar H}_3\\
    q_3 =& \frac{Lx/h-{\cal Q}_1}{x^2-1}&-H_2+{H}_1+{\bar H}_{3}-{\bar H}_2\\
    q_4 =& \frac{Lx/h-{\cal Q}_1}{x^2-1}&-H_3+{H}_2+{\bar H}_{2}-{\bar H}_1\\
    q_1 =& \frac{Lx/h+{\cal Q}_2x}{x^2-1}&+H_4+{H}_{\bar 4}-{H}_{3}-{\bar H}_1\\
    q_5 =& &+H_4-{H}_{\bar 4}+{\bar H}_{4}-{\bar H}_{\bar 4}\\
    q_a =& -q_{11-a}\;\;,&a=6,\dots,10\;,
    \eea
    \la{eq:p}
    \eeqa
    where the resolvents $H_a$ have the form
    \beqa
    &&\nn H_a(x)=\sum_j\frac{x^2}{x^2-1}\frac{1}{x-x_{a,j}}\;\;,\;\;\bar
    H_a(x)=H_a(1/x)\;.
    \eeqa
    In these terms the Bethe equations \eq{ABA} are
    equivalent to the condition that the two eigenvalues
    of the monodromy matrix
    are equal along the branch cut
    \beq
    q_i(x+i0)-q_j(x-i0)=2\pi n\;\;,\;\;x\in{\cal C}\;.
    \eeq

    We can now simplify \eq{eqW} for strong coupling.
    First of all we notice that the shift operator $D$
    becomes a formal expansion parameter. Then we use
    \beqa
    \nn\frac{Q_1^- B^{(+)+}R^{(+)-}}{Q_1^+ B^{(-)+}R^{(-)-}}
    &\simeq&\exp\[-i\(H_1-H_4-H_{\bar 4}+\bar H_1+\bar H_4+\bar H_{\bar 4}\)\]
    \\
    \nn\frac{Q_3^{+}Q_2^{--}  R^{(+)-}}{e^{\half i \cQ_1}Q_3^-Q_2  R^{(-)-}}
    &\simeq&
    \exp\[-i\(\frac{{\cal Q}_1+x {\cal Q}_2}{x^2-1}+H_2-H_3-\bar H_3+\bar H_2-H_4-H_{\bar 4}\)\]
    \\
    \nn\frac{Q_1^{+}Q_2^{--}  R^{(+)-}}{e^{\half i \cQ_1}Q_1^-Q_2  R^{(-)-}}
    &\simeq&
    \exp\[-i\(\frac{{\cal Q}_1+x {\cal Q}_2}{x^2-1}+H_2-H_1-\bar H_1+\bar H_2-H_4-H_{\bar 4}\)\]
    \\
    \frac{Q_3^+}{Q_3^-}
    &\simeq&
    \exp\[+i\(H_3+\bar H_3\)\]\;.
    \eeqa
    The generating functional \eq{eqW} becomes
    \beq\la{Wstr}
    {\cal W}=
    \frac{(1-\lambda_1d)(1-\lambda_2d)}{(1-\lambda_3d)(1-\lambda_4d)}\;,
    \eeq
    where we have redefined the formal expansion parameter in the following way
    \beq
    d=\exp\[i\(\frac{Lx/h+x{\cal Q}_2}{x^2-1}+H_4+H_{\bar 4}-\bar H_1+\bar H_3\)\]D\;.
    \eeq
    Expanding the generating function \eq{Wstr} we get
    \beqa
    T_{1,s}&=&\frac{ {\lambda_4}^{s-1} ( {\lambda_4}- {\lambda_1})
       ( {\lambda_4}- {\lambda_2})- {\lambda_3}^{s-1} ( {\lambda_3}- {\lambda_1})
       ( {\lambda_3}- {\lambda_2})}{ {\lambda_4}- {\lambda_3}}\(\frac{d}{D}\)^s\\
    T_{a,1}&=&(-1)^a\frac{ {\lambda_1}^{a-1} ( {\lambda_1}- {\lambda_3})
       ( {\lambda_1}- {\lambda_4})- {\lambda_2}^{a-1} ( {\lambda_2}- {\lambda_3})
       ( {\lambda_2}- {\lambda_4})}{ {\lambda_1}- {\lambda_2}}\(\frac{d}{D}\)^a\;.\nn
    \eeqa
    It is now straightforward to compute ${\bf Y}_{\fp_a}$ and ${\bf Y}_{\fb_s}$ from \eq{YbtoT}.
    Note that the factors $\(\frac{d}{D}\)^s$ and $\(\frac{d}{D}\)^a$ are irrelevant here
    and thus ${\bf Y}_{\fp_a}$ are rational functions of $\lambda_a$ only!

    Moreover, using the relation
    \beqa
    \(\frac{x^-}{x^+}\)^L\Phi_4(u)
    &\simeq&\exp\[-i\(\frac{xL/h+x{\cal Q}_2}{x^2-1}+2H_{\bar 4}-\bar H_4+\bar H_{\bar 4}-\bar H_1+\bar H_3\)\]\;,
    \eeqa
    and the same relation with $4$ and $\bar 4$ exchanged, we obtain expressions for the massive nodes:

    \beqa
    Y_{\fo_a}&=&\(-\lambda_5\)^{-\omega_a}
    \frac{ {\lambda_1}^{a-1} ( {\lambda_1}- {\lambda_3})
       ( {\lambda_1}- {\lambda_4})- {\lambda_2}^{a-1} ( {\lambda_2}- {\lambda_3})
       ( {\lambda_2}- {\lambda_4})}{ {\lambda_1}- {\lambda_2}}\;,\nn\\
    Y_{\fO_a}&=&\(-\lambda_5\)^{+\omega_a}
    \frac{ {\lambda_1}^{a-1} ( {\lambda_1}- {\lambda_3})
       ( {\lambda_1}- {\lambda_4})- {\lambda_2}^{a-1} ( {\lambda_2}- {\lambda_3})
       ( {\lambda_2}- {\lambda_4})}{ {\lambda_1}- {\lambda_2}}\;,
    \eeqa
    which are again written solely in terms of the eigenvalues of the
    classical monodromy matrix! Here, we have introduced $\omega_a$, which is defined to be 1 for odd $a$ and zero for even $a$.

    \section{TBA equations for ${\rm AdS}_4/{\rm CFT}_3$}
    In this section we derive the Thermodynamic Bethe ansatz equations for ${\rm AdS}_4/{\rm CFT}_3$. Let us first describe
    the general form of the TBA method~\cite{Zamolodchikov:1989cf}
     (see a nice introductory
    paper \cite{Arutyunov:2007tc}). We start with an integrable quantum
    field theory in 1+1 dimensions, on a circle of circumference $L$. The partition
    function of this theory at temperature $1/R$ is
    \be
    \label{ZDef}
        Z(L,R) = \sum_k \e^{-RE_k(L)},
    \ee
    and in the limit $R \to \infty, \ R \gg L$ we have
    \be
    \label{ZviaGS}
        Z(L,R) \sim \e^{-RE_0(L)} ,
    \ee
    where $E_0(L)$ is the ground state energy. Denoting by $\phi$ and $\psi$ the bosonic and fermionic fields, respectively, we can write the partition function as a functional integral
    \be
        Z(L,R) = \int \mathcal D\phi\mathcal D\psi \e^{-S_E}
    \ee
    where $S_E$ is the theory's Euclidean action. In this integral, fermionic fields are periodic (resp. antiperiodic) in space (resp. time), while bosonic fields are periodic in both space and time:
    \be
    \begin{array}{rclrcl}
        \psi(x+L,t) &=& \psi(x,t), \ \psi(x,t+R) &=& -\psi(x,t)\\
        \phi(x+L,t) &=& \phi(x,t), \ \phi(x,t+R) &=& \phi(x,t).\\
    \end{array}
    \ee
    Using this representation of the partition function, one can relate it
    to the Witten index of the ``mirror" theory in volume $R$:
    \be
    \label{WittenInd}
        W(R,L) = \sum_k (-1)^{F} \e^{-L E^{\mir}_k(R)} = \sum_k \e^{-RE_k(L)} = Z(L,R)\;.
    \ee
    The mirror theory is obtained from the original one by a double Wick rotation, and $F$  in~\eq{WittenInd} is 1 for fermionic states and 0 otherwise.
    Introducing the mirror bulk free energy $\mathcal F^{\rm mir}(L)$, defined by the mirror theory's Witten index at temperature $1/L$,
    \be
    \label{ZamFreeEnDef}
        -RL \mathcal{F}^{\rm mir}(L) = \ln W(R,L),
    \ee
    we see that the finite volume ground state energy is related to the infinite volume mirror free energy:
    \be
    \label{ZamEandF}
    E_0(L) = L \mathcal F^{\rm mir}(L).
    \ee
    The mirror theory's infinite volume spectrum is described by the ABA equations, which allow one to find $\mathcal F^{\rm mir}(L)$ and then the original theory's ground state energy.

    To compute $\mathcal F^{\rm mir}$ it is essential to know the structure of the solutions of infinite volume mirror ABA equations. For numerous theories (see~\cite{Zamolodchikov:1991et}), so-called string hypotheses have been formulated, which describe the complexes Bethe roots form in the infinite volume limit (simplest of those complexes are strings of roots). We will use indices $A,B,...$ to label the complexes, and denote the energy and momentum of a complex by, respectively, $ip_A^*$ and $i\eps^*_A$, to underline that the mirror theory is obtained from the physical one by a double Wick rotation.

    Multiplying the Bethe equations for all roots in a complex, one obtains equations for the density $\rho_A (u)$ of complexes, with $u \in \mathbb R$ being the center of the complex. Those equations have the form
    \be
        \label{TD_BAE_gen_mir}
       \bar\rho_A(u)+ \rho_A(u)=
       \frac{i}{2\pi} \frac{d \eps^*_A(u)}{du} -K_{BA}(v,u)*\rho_B(v)\,\,
    \ee
    where $\bar\rho$ is the density of holes, $K(v,u) * f(v) \equiv \int^{+\infty}_{-\infty} dv K(v,u) f(v)$ and summation over $B$ is assumed.
    Also, we use the normalization
    \be
        \int\limits_{-\infty}^{\infty} du \rho_A(u) =
        \frac{\text{total number of complexes of type } A}{R}.
    \ee

    The free energy is given by the minimal value of a functional of the densities
    \be
    \label{CalFmir}
        \mathcal{F}^{\rm mir}(L)= \min \,\,\, \sum_{A}  \int_{-\infty}^\infty\,  du  \left(
        (Lip_A^* + h_A) \rho_A-          \left[ \rho_A
        \log\left(1+\frac{\bar\rho_A}{\rho_A}\right)+  \bar\rho_A
        \log\left(1+\frac{\rho_A}{\bar\rho_A}\right)\right] \right),
    \ee
    with constraints~\eq{TD_BAE_gen_mir} on the densities. Here,
    $h_A \equiv \log[(-1)^{N_A}]$ , where $N_A$ is the number of fermionic Bethe roots in the complex $A$. Minimization of this functional gives the TBA equations
    \beqa
    \label{TBAfirst_mir}
        \log {\cal Y}_A(u) &=& K_{AB}(u,v)*\log[1+1/{\cal Y}_B(v)]+ iLp_A^* + h_A,
    \eeqa
    where $\mathcal{Y}_A \equiv \frac{\bar \rho_A}{\rho_A}$ and $K(u,v) * f(v) \equiv \int dv K(u,v) f(v)$. Lastly, the free energy can be expressed in terms of a solution of TBA equations:
    \be
    \label{FmirRes}
        \mathcal{F}^{\rm mir} (L) = \sum_{A} \int  \frac{du}{2\pi i} \frac{d\eps^*_A}{du}
        \,\log\left(1+1/\mathcal{Y}_A(u) \right),
    \ee
    From the free energy, one can compute the ground state energy of the physical theory via~\eq{ZamEandF}. In addition, the TBA equations can be modified in such a way that their solutions provide also energies of certain excited states in finite volume.

    \subsection{Ground state TBA equations for ${\rm AdS}_4/{\rm CFT}_3$.}
    We first present, as a conjecture, the ABA equations for the mirror of ${\rm AdS}_4/{\rm CFT}_3$ theory. Like the physical Bethe equations~\cite{GV2}, those equations involve Bethe roots
    $u_1,u_2,u_3,u_4,u_{\bar 4}$,
    with all roots except $u_2$ being fermionic. The only roots which carry energy or momentum are $u_4$ and $u_{\bar 4}$. For a single root, we denote energy by $ip_1^*$, and momentum by $i\eps_1^*$, where
    \be
    \label{pepsDef}
        p_1^* = \frac{1}{i}\log\frac{x^+}{x^-}, \ \ \
        \epsilon_1^* =
        \half + h(\lambda) \left( \frac{i}{x^+} - \frac{i}{x^-}\right).
    \ee
Here, and everywhere in section 3 unless otherwise stated, we use the mirror branch of the function $x(u)$. Note that $p_1^*$ (resp. $\eps_1^*$), evaluated in physical instead of mirror kinematics, coincides with the momentum (resp. energy) of a single Bethe root in the physical theory~\cite{GV2}. This is in accordance with the fact that the mirror theory is obtained from the physical one by a double Wick rotation. The momentum/energy in mirror and physical ${\rm AdS}_5/{\rm CFT}_4$ are related in a similar way.

    The mirror Bethe equations we propose are written in terms of the functions $B_l, R_l, S_l, Q_l$, which were introduced in section 2 (note that $x$ in them should now be understood as $x^{\mir}$). The equations for $u_1, u_2$ and $u_3$ are:
    \be
        1
        =
        \left.
        \frac{Q^+_2 B_{4}^{(-)} B_{\bar4}^{(-)}}{Q^-_2 B_{4}^{(+)} B_{\bar4}^{(+)} }
        \right|_{u_{1,k}},
        \ \
        -1 =
        \left.
        \frac{Q^{--}_2 Q^+_1 Q^+_3}{Q^{++}_2 Q^-_1 Q^-_3}
        \right|_{u_{2,k}},
        \ \
        1 =
        \left.
        \frac{Q^+_2 R_{4}^{(-)} R_{\bar4}^{(-)}}{Q^-_2 R_{4}^{(+)} R_{\bar4}^{(+)} }
        \right|_{u_{3,k}}
    \ee
The r.h.s. of the equations for $u_1$ and $u_3$
is not always unimodular, because $B^{(\pm)}_l (u)$ and $R^{(\pm)}_l (u)$ have cuts on the real axis. However, in the thermodynamic limit (see below) the single fermion roots $u_1, u_3$ are distributed \cite{Arutyunov:2009ur} within the interval $-2h<u_1<2h$, $-2h<u_3<2h$, and unimodularity of the r.h.s then follows. Note that no conditions have to be imposed on the $u_1, u_3$ roots which are parts of pyramid complexes $\fp_n$, as the terms containing cuts cancel during fusion of Bethe equations. This can be seen from the fact that the kernels $\mathcal{K}(u,v)$ in TBA equations (see below) for interactions involving pyramids are real for real $u,v$ and have no cuts on the real axis.

The equations for momentum-carrying roots are:
    \be
        -1 =
        \e^{R\eps^*_1(u_{4,k})}
        \left(
        \frac{B^+_1 R^+_3 Q^{++}_4 R^{-(-)}_{4} R^{-(-)}_{\bar4}}
              {B^-_1 R^-_3 Q^{--}_4 R^{+(+)}_{4} R^{+(+)}_{\bar4}}
        \right)
        S^{}_{4} S^{}_{\bar4}
        \left(\frac{x_{4,k}^+}{x_{4,k}^-}\right)^{\frac{K_{1}-K_{3}}{2}}
        \prod_{j=1}^{K_4}\sqrt{\frac{x_{4,j}^+}{x_{4,j}^-}}
        \prod_{j=1}^{K_{\bar 4}}\sqrt{\frac{x_{\bar 4,j}^+}{x_{\bar 4,j}^-}}
    \ee
for $u=u_{4,k}$, and for $u=u_{\bar 4,k}$ we have
    \be
        -1 =
        \e^{R\eps^*_1(u_{\bar4,k})}
        \left(
        \frac{B^+_1 R^+_3 Q^{++}_{\bar4} R^{-(-)}_{\bar4} R^{-(-)}_{4}}
              {B^-_1 R^-_3 Q^{--}_{\bar4} R^{+(+)}_{\bar4} R^{+(+)}_{4}}
        \right)
        S^{}_{\bar4} S^{}_{4}
        \left(\frac{x_{\bar 4,k}^+}{x_{\bar 4,k}^-}\right)^{\frac{K_{1}-K_{3}}{2}}
        \prod_{j=1}^{K_4}\sqrt{\frac{x_{4,j}^+}{x_{4,j}^-}}
        \prod_{j=1}^{K_{\bar 4}}\sqrt{\frac{x_{\bar 4,j}^+}{x_{\bar 4,j}^-}}.
    \ee
Note that the combination
\beq
    S^{\mir}_l (u)\equiv\prod_{j=1}^{K_l} \sigma^{\mir} (x(u), x_{l,j})
    =\prod_{j=1}^{K_l} \sqrt{\frac{x_{l,j}^+x^-}{x_{l,j}^-x^+}}\sigma_{BES} (x(u), x_{l,j})
    =S_l(u)\prod_{j=1}^{K_l} \sqrt{\frac{x_{l,j}^+x^-}{x_{l,j}^-x^+}}
\eeq
is a unimodular function (see \cite{Gromov:2009bc}, \cite{Arutyunov:2009kf}).
By $\sigma_{BES} (x(u), x_{l,j})$ we denote the usual Beisert-Eden-Staudacher dressing kernel
analytically continued from $\IM u>i/2$ between the branch points $u=\pm 2h+i/2$.\footnote{``Physical" choice of the branch
corresponds to analytical continuation to the plane with the cut $[-2h+i/2,2h+i/2]$. In the ``mirror"
kinematics all cuts should go through infinity.}
    The above equations are similar to the Bethe equations for
    physical ABA \eq{ABA}. The difference is in the choice of the mirror branch of
    $x(u)$, interchange of the energy and momentum (with multiplication by $i$) and various factors of $\sqrt{x^+/x^-}$,
    tuned in such a way that the right-hand sides are unimodular functions.
    This prescription is based on the corresponding conjecture in the ${\rm AdS}_5/{\rm CFT}_4$ case \cite{Arutyunov:2007tc}.

    In the thermodynamic limit, solutions of the above ABA are described by complexes of Bethe roots.
    Among those complexes are $\figb_n, \figp_n, \figf, \figF$, which are the same complexes as
    in the mirror ${\rm AdS}_5/{\rm CFT}_4$ (see~\cite{Arutyunov:2009zu}). In addition, the momentum-carrying
    roots $u_4$ and $u_{\bar 4}$ form two new types of complexes, which we call Odd-Even and Even-Odd. They were recently considered in \cite{Saleur:2009bf}.
    Those complexes are real-centered strings of alternating $u_4$ and $u_{\bar4}$ roots,
    adjacent roots being spaced by $i$.  In the Odd-Even complex, the lowest root of the string
    on the complex plane is $u_{4}$, while in the Even-Odd complex, the lowest root is $u_{\bar4}$ (see Fig. 2). The list of all complexes is given in the table below.
    \begin{equation*}
    \begin{array}{rccccc}
    \hline
        \figb_n :&\text{string of roots}&:&u_{2}=u+ij, & j=-\frac{n-2}{2},\dots,\frac{n-2}{2}
    \\
    \hline
        :&&:&u_{3}=u+ij, & j=-\frac{n-1}{2},\dots,\frac{n-1}{2}
    \\
        \figp_n :&\text{pyramid}&:& u_{2}=u+ij, & j=-\frac{n-2}{2},\dots,\frac{n-2}{2}
    \\
        :&&:&u_{1}=u+ij, & j=-\frac{n-3}{2},\dots,\frac{n-3}{2}
    \\
    \hline
        \figf :&\text{single fermion root}&:& u_{1}=u
    \\
    \hline
        \figF :&\text{single fermion root}&:& u_{3}=u
    \\
    \hline
        \fo_n :&\text{Odd-Even complex}&:&
        u_4 = u+ij \ \ \text{when}\ \ \theta^E_{jn}=1, \ \ \ \ \ \ &
        j=-\frac{n-1}{2}, \dots,\frac{n-1}{2}

    \\
        :&&:&u_{\bar4} = u+ij\ \ \text{when}\ \ \theta^O_{jn}=1 \ \ \ \ \ \ &
    \\
    \hline
        \fO_n :&\text{Even-Odd complex}&:&
        u_4 = u+ij \ \ \text{when}\ \ \theta^O_{jn}=1, \ \ \ \ \ \ &
        j=-\frac{n-1}{2}, \dots,\frac{n-1}{2}

    \\
        :&&:&u_{\bar4} = u+ij\ \ \text{when}\ \ \theta^E_{jn}=1 \ \ \ \ \ \ &
    \\
    \hline
    \end{array}
    \end{equation*}

    \FIGURE[ht]{\la{Fig:AltCompl}
    \begin{tabular}{lr}
    \includegraphics[scale=0.6]{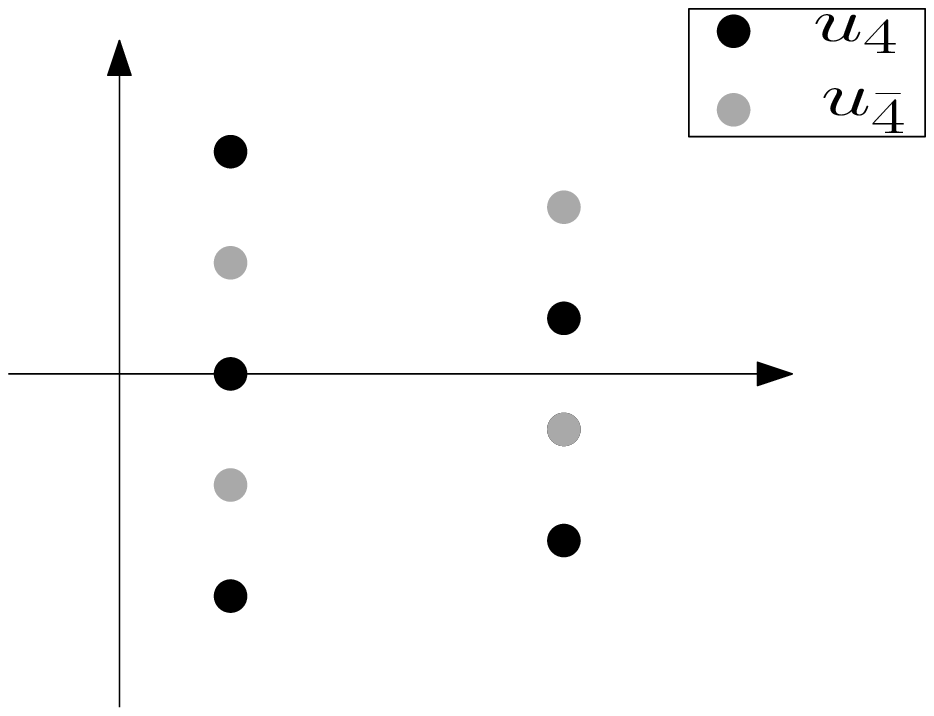}
    &
    \includegraphics[scale=0.6]{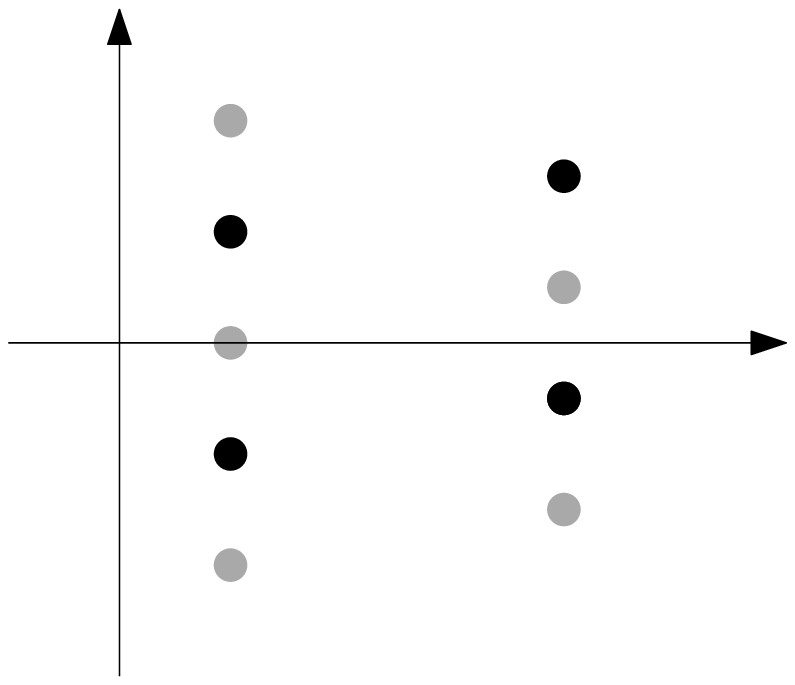}
    \\
    {Odd-Even complexes $\fo_5$ and $\fo_4$}
    &
    {Even-Odd complexes $\fO_5$ and $\fO_4$}
    \end{tabular}
    \caption{Strings of alternating roots on the $\{{\rm Re}\,  u, {\rm Im}\, u\}$ plane. Black circles denote $u_4$ roots, gray circles denote $u_{\bar4}$ roots. Vertical spacing between roots is $i$.
    }
    }

    Here, $u \in \mathbb R$ denotes the center of a complex, notation $j=-\frac{n-1}{2},\dots,\frac{n-1}{2}$ means that $j$ takes the values $-\frac{n-1}{2},-\frac{n-3}{2}, \dots,\frac{n-3}{2},\frac{n-1}{2}$, and $\theta$'s were defined in
    \eq{eqtheta}.


    The energy (in our notation $ip^*_A$, with index $A$ taking the values $\{ \figb_n, \figf, \figF, \figp_n, \fo_n, \fO_n\}$) which corresponds to a complex is the sum of energies of the roots in a complex, and the same is true for momentum. We have $p^*_{\fo_n} = p^*_{\fO_n} = p^*_n$, $\epsilon^*_{\fo_n} = \epsilon^*_{\fO_n} = \eps^*_n$, where
    \be
        p^*_n (u) \equiv \frac{1}{i}\log\frac{x^{[+n]}}{x^{[-n]}}, \ \
        \eps^*_n (u) \equiv
        \frac{n}{2} + h(\lambda) \left( \frac{i}{x^{[+n]}} - \frac{i}{x^{[-n]}}\right),
    \ee
    while for other complexes, $p^*_A$ and $\epsilon^*_A$ are zero. Note also that the only complexes with odd number of fermion roots are those denoted by ${\figf},\;{\figF}$, ${\fo_{2n-1}}$ and ${\fO_{2n-1}}$. Hence in our case the quantity $h_A$ in ~\eq{TBAfirst_mir}
has to be $\log(-1)$ for these complexes and $\log(+1)$ otherwise.

    Applying the fusion procedure to the mirror ABA equations\footnote{an important assumption here is monotonicity}, we find that in our case, kernels $K_{AB}$ entering~\eq{TBAfirst_mir} are given by the table below.
    \beqa
    \nn
    {
    \footnotesize
    \begin{array}{|l@{}||l@{}|l|l|l|l|l|}\hline
    A\backslash B &\figb_m&\figf&\figF&\figp_m&\fo_m&\fO_m\\ \hline\hline
        \figb_n
            & +K_{n-1,m-1} & -K_{n-1} & +K_{n-1} & 0 & 0 &0\\ \hline
        \figf
            & -K_{m-1} & 0 & 0&+K_{m-1}
            & -\mathcal B^{(01)}_{1m} &-\mathcal B^{(01)}_{1m}\\ \hline
        \figF
            & -K_{m-1} & 0 & 0 &+K_{m-1}
            & -\mathcal R^{(01)}_{1m} &-\mathcal R^{(01)}_{1m}\\ \hline
        \figp_n
            &0& -K_{n-1}&+K_{n-1}&+K_{n-1,m-1}
            &
            -\mathcal R^{(01)}_{nm}-\mathcal B^{(01)}_{n-2,m}
            &-\mathcal R^{(01)}_{nm}-\mathcal B^{(01)}_{n-2,m}
            \\ \hline
        \fo_n
            & 0 & \mathcal B^{(10)}_{n1} & -\mathcal R^{(10)}_{n1} &
    -\mathcal{R}^{(10)}_{nm}
            -\mathcal{B}^{(10)}_{n,m-2}
            & \bal
            -{\cal T}^{\|}_{nm}
            \eal
          & \bal
    -{\cal T}_{nm}^{\bot}
          \eal
            \\ \hline
        \fO_n
            & 0 & \mathcal B^{(10)}_{n1} & -\mathcal R^{(10)}_{n1} &
    -\mathcal{R}^{(10)}_{nm}
            -\mathcal{B}^{(10)}_{n,m-2}
            & \bal
    -{\cal T}_{nm}^{\bot}
            \eal
          & \bal
    -{\cal T}_{nm}^{\|}      \eal
            \\ \hline
    \end{array}}
    &&
    \eeqa
    Some of those kernels are the same as in the ${\rm AdS}_5/{\rm CFT}_4$ case~\cite{Gromov:2009bc}, and we list them in Appendix A. The new kernels are
    \beqa
    {\cal T}_{nm}^{\|}&\equiv&\mathcal{B}^{(11)}_{nm}+\tilde{\cal S}_{nm}-K_{nm}^{\|},\\
    {\cal T}_{nm}^{\bot}&\equiv&        \mathcal{B}^{(11)}_{nm}
            +\tilde{\cal S}_{nm}-K_{nm}^{\bot},
    \eeqa
    where
    \beqa
        K^{\|}_{nm}(u,v) &\equiv&
        \sum_{l=-\frac{n-1}{2}}^\frac{n-1}{2}\sum_{k=-\frac{m-1}{2}}^\frac{m-1}{2}
        K_2(u-v+i(l-k))\(\theta^{\rm E}_{ln}\theta^{\rm O}_{km}+\theta^{\rm O}_{ln}\theta^{\rm E}_{km}\),\\
        K^{\bot}_{nm}(u,v) &\equiv&
        \sum_{l=-\frac{n-1}{2}}^\frac{n-1}{2}\sum_{k=-\frac{m-1}{2}}^\frac{m-1}{2}
        K_2(u-v+i(l-k))\(\theta^{\rm E}_{ln}\theta^{\rm E}_{km}+\theta^{\rm O}_{ln}\theta^{\rm O}_{km}\).
    \eeqa
    %

    Let us also introduce the functions $Y_A$, which will turn out to be the functions which enter the Y-system:
    \begin{eqnarray}
        \left\{{Y}_{\figb_{ n}} ,Y_{\figf},\frac{1}{Y_{\figF}},\frac{1}{Y_{\figp_{ n}}},
        \frac{1}{Y_{\fo_{ n}}},
        \frac{1}{Y_{\fO_{ n}}  }\right\}
        &\equiv&
        \left\{\mathcal{Y}_{\figb_{n}},\mathcal{Y}_{\figf_{}},\mathcal{Y}_{\figF_{}},
        \mathcal{Y}_{\figp_{n}},
        \mathcal{Y}_{\fo_{ n}}, \mathcal{Y}_{\fO_{ n}}\right\}
    \end{eqnarray}
    (We recall that $\mathcal Y_A \equiv \frac{\bar\rho_A}{\rho_A}$.)

    \bigskip

    We can write the TBA equations for the ground state in the following way:
    \beqa
    \label{YFmir}
        \log Y_{\figF}&=&
        +K_{m-1}*\log\frac{1+1/Y_{\figb_{m}}}{1+Y_{\figp_{m}}}
        +\mathcal{ {R}}_{1m}^{(01)}*\log(1+Y_{\fo_m})(1+Y_{\fO_m})
        + i\pi
    \\
    \label{Yfmir}
        \log Y_{\figf }&=&
        -K_{m-1}*\log\frac{1+1/Y_{\figb_{m}}}{1+Y_{\figp_{m}}}
        -\mathcal{ {B}}_{1m}^{(01)}*\log(1+Y_{\fo_m})(1+Y_{\fO_m})
        -i\pi
    \\
    \label{Ypmir}
        \log Y_{{\figp}_{n}}&=&-K_{n-1,m-1}*\log(1+Y_{{\figp}_{m}})
        -K_{n-1}\cut\log({1+Y_{\figF}})
    \\ \nn
        &+&
         \left({\cal  {R}}^{(01)}_{nm}+{\cal
         {B}}^{(01)}_{n-2,m}\right)*\log(1+Y_{\fo_m})(1+Y_{\fO_m})
    \\
    \label{Ybmir}
        \log Y_{{\figb}_{n}}&=&K_{n-1,m-1}*\log(1+1/Y_{{\figb}_{m}})
        +K_{n-1}\cut\log({1+Y_{\figF}})
    \\
    \label{Yomir}
        \log Y_{{\fo}_{n}}&=&
        -iLp^*_n
        + {\cal T}_{nm}^{\|}
        * \log(1+Y_{\fo_m})
        + {\cal T}_{nm}^{\bot}
        * \log(1+Y_{\fO_m})+i\pi n
    \\ \nn
        &+&
        {\cal  {R}}_{n1}^{(10)}\cut\log(1+Y_{\figF})
        +\left({\cal  {R}}_{nm}^{(10)}+{\cal  {B}}_{n,m-2}^{(10)}\right)
        * \log(1+Y_{\figp_m})
    \\
    \label{YOmir}
        \log Y_{{\fO}_{n}}&=&
        -iLp^*_n
        + {\cal T}_{nm}^{\|}
        * \log(1+Y_{\fO_m})
        + {\cal T}_{nm}^{\bot}
        * \log(1+Y_{\fo_m})+i\pi n
    \\ \nn
        &+&{\cal  {R}}_{n1}^{(10)}\cut\log(1+Y_{\figF})
        +\left({\cal  {R}}_{nm}^{(10)}+{\cal  {B}}_{n,m-2}^{(10)}\right)
        * \log(1+Y_{\figp_m})
    \eeqa
    where $*$ denotes integration over the second variable, as in~\eq{TBAfirst_mir}. Summation over the repeated index $m$ is assumed  with $m\ge 2$ for $\figp_{m}$
    and $\figb_{m}$, and $m\ge1$ for $\fo_m,\fO_m$.

    Range of integration for fermions is limited to $-2h<u<2h$.
   Notice that from \eq{YFmir} and \eq{Yfmir} we can see that $\frac{1}{Y_{\figf}}$
is the analytical continuation of $Y_{\figF}$   across the cut
   $u\in (-\infty,-2h)\cup(2h,+\infty)$.
       For the convolutions with fermions we introduce the convolutions
       $\cut$ which should be understood in the sense of a B-cycle (see \cite{Gromov:2009bc}), e.g.
\beqa
&&K_{n-1}\cut\log({1+Y_{\figF}})\equiv\int_{-2h}^{2h} dv
K_{n-1}\log\frac{1+Y_{\figF}}{1+1/Y_{\figf}}\;,\\
&&{\cal R}^{(n0)}\cut\log(1+Y_{\figF})\equiv\int_{-2h}^{2h} dv \left[ {\cal R}^{(n0)}\log(1+Y_{\figF})-{\cal B}^{(n0)}\log(1+1/Y_{\figf})
\right] \nn\;.
\eeqa

    Remarkably, the combination
    $\mathcal{B}^{(11)}_{nm}+\tilde{\cal S}_{nm}$, which is part of the kernels ${\cal T}_{nm}^{\bot}$ and ${\cal T}_{nm}^{\|}$, has only two branch cuts for each of the variables $u$ and $v$. This follows from the integral representation
\beqa
\label{SBintrep}
    &&{\cal \tilde S}_{nm}(u,v)+{\cal B}_{nm}^{(11)}(u,v)
    =\\
    &&-\sum_{a=1}^{\infty}\int
    \left[{\cal B}^{(10)}_{n1}\left(u,w+ia/2\right){\cal B}_{1m}^{(01)}\left(w-ia/2,v\right)
     + {\cal B}^{(10)}_{n1}\left(u,w-ia/2\right){\cal B}_{1m}^{(01)}\left(w+ia/2,v\right)\right]dw \nn,
\eeqa
    which can be derived using the results obtained in \cite{Gromov:2009bc,Arutyunov:2009kf} (see Appendix A).
    As a consequence, the functions $Y_{\fo_n}(u)$
    and $Y_{\fO_n}(u)$ (see \eq{Yomir}, \eq{YOmir}) should not have branch cuts for $-in/2<\IM u<in/2$.

In the next section we will establish a relation between the above equations and the ${\rm AdS}_4/{\rm CFT}_3$ Y-system.
    \subsection{Y-system from TBA equations}
    In this section we show that solutions of ground state TBA equations satisfy the ${\rm AdS}_4/{\rm CFT}_3$ Y-system \eq{YoOeq1}--\eq{YoOeq3} described in the introduction. This derivation of the Y-system is similar to the ${\rm AdS}_5/{\rm CFT}_4$ case~\cite{Gromov:2009bc}. First, we identify the $Y_A$ functions in TBA equations with the Y-system functions $Y_{a,s}$. We set
    \be
        \left\{ Y_{\figb_n}, Y_{\figf}, Y_{\figF}, Y_{\figp_n} \right\} =
        \left\{ Y_{1,n}, Y_{2,2}, Y_{1,1}, Y_{n,1} \right\}.
    \ee

    \bigskip

    Let us introduce the discrete Laplacian operator
    \be
        \Delta K_n(u)\equiv K_n(u+i/2-i0)+K_n(u-i/2+i0)-K_{n+1}(u)-K_{n-1}(u)\nn.
    \ee
    Following~\cite{Gromov:2009bc}, we apply this operator to the l.h.s. of the TBA equations, acting on the free index $n$ and the free variable $u$. The action of this Laplacian on some of our kernels has been computed in~\cite{Gromov:2009bc}:
    \beqa
        \Delta K_n(u)&=&\delta_{n,1}\delta(u)
    \\
        \Delta K_{nm}(v-u)&=&\Delta{\cal R}^{(11)}_{nm}(v,u)=
        \delta_{n,m+1}\delta(v-u)+\delta_{n,m-1}\delta(v-u)
    \\
        \Delta{\cal R}^{(01)}_{nm}(v,u)&=&
        \Delta{\cal R}^{(10)}_{nm}(v,u)=\delta_{n,m}\delta(v-u)
    \\
        \Delta{\cal B}_{nm} &=& 0, \ \ \ \Delta\tilde{\cal S}_{nm} = 0\;.
    \eeqa
    The new kernels
    ${\cal T}_{nm}^{\|}$ and ${\cal T}_{nm}^{\bot}$ satisfy relations of a new type, which are not written in terms of the Laplacian:
    \beqa
    \la{newlap}\!\!\!\!\!\!\!\!\!\!\!\!\!\!\!\!&&{\cal T}_{nm}^{\bot}(u\!+\!\tfrac{i-i0}{2},v)\!+\!
    {\cal T}_{nm}^{\|}(u\!-\!\tfrac{i-i0}{2},v)\!-\!
    {\cal T}_{n+1,m}^{\|}(u,v)\!-\!
    {\cal T}_{n-1,m}^{\bot}(u,v)\!=\!-\delta_{n,m-1}\delta(u\!-\!v)\\
    \nn\!\!\!\!\!\!\!\!\!\!\!\!\!\!\!\!&&{\cal T}_{nm}^{\|}(u\!+\!\tfrac{i-i0}{2},v)\!+\!
    {\cal T}_{nm}^{\bot}(u\!-\!\tfrac{i-i0}{2},v)\!-\!
    {\cal T}_{n+1,m}^{\bot}(u,v)\!-\!
    {\cal T}_{n-1,m}^{\|}(u,v)\!=\!-\delta_{n-1,m}\delta(u\!-\!v)\;.
    \eeqa
    Using those identities, we obtain from the TBA equations a set of simpler equations
    for the functions $Y_A$. This closely follows~\cite{Gromov:2009bc}. For example, applying the Laplacian to the l.h.s of~\eq{Ybmir}, we get
    \be
        \log\frac{Y_{{\figb}_{n}}^+ Y_{{\figb}_{n}}^-}{\Yb{n+1} \Yb{n-1}}
        =\log(1+1/Y_{{\figb}_{n+1}})(1+1/Y_{{\figb}_{n-1}}) \,\, , \,\, n>2
    \ee
    or, equivalently,
    \be
        \log{Y_{{\figb}_{n}}^+Y_{{\figb}_{n}}^-}
        =\log(1+Y_{{\figb}_{n+1}})(1+Y_{{\figb}_{n-1}}) \,\, , \,\, n>2.
    \ee
    For $n=2$ we obtain
    \be
        \log{Y_{{\figb}_{2}}^+Y_{{\figb}_{2}}^-}
        =\log\frac{(1+Y_{\figF})(1+Y_{{\figb}_{3}})}{1+1/Y_{\figf}}.
    \ee
    Equations~\eq{YFmir} and~\eq{Ypmir} can be treated in a similar way. We get an equation for $\YF$
    \be
        \log{Y_{\figF}^+ Y_{\figF}^-}=
        \log\frac{(1+Y_{\figb_2})(1+Y_{\fo_1})(1+Y_{\fO_1})}{1+1/Y_{\figp_2}},
    \ee
    and also equations for $Y_{{\fp}_{n}}$:
    \be
        \log\frac{Y_{{\figp}_{n}}^+Y_{{\figp}_{n}}^-}{Y_{{\figp}_{n+1}}Y_{{\figp}_{n-1}}}
        =\log\frac{(1+Y_{\fo_n})(1+Y_{\fO_n})}{(1+Y_{{\figp}_{n+1}})(1+Y_{{\figp}_{n-1}})} \,\, , \,\, n>2
    \ee

    \beqa
    \label{YPyr2init_mir}
        \log \frac {Y_{\figp_2}^+ Y_{\figp_2}^-} { Y_{\figp_3}}
        &=&
        \log\frac {(1+Y_{\figf})(1+Y_{\fo_2})(1+Y_{\fO_2}) Y_{\figF}}
         {(1+Y_{\figp_3})(1+Y_{\figF})}
    \\ \nn
        &-& \log Y_{\figF}Y_{\figf}
        +
          \sum_m (\mathcal{{R}}^{(01)}_{1m}-\mathcal{{B}}^{(01)}_{1m})
          *\log(1+Y_{\fo_m})(1+Y_{\fO_m})
    \eeqa
    Moreover, adding up Eqs.~\eq{YFmir},~\eq{Yfmir} we find that
    \be
    \label{TBA_fin_f_mir}
        \log Y_{\figF}Y_{\figf} =
           \sum_m (\mathcal{{R}}^{(01)}_{1m}-\mathcal{{B}}^{(01)}_{1m})
           *\log(1+Y_{\fo_m})(1+Y_{\fO_m}).
    \ee
    Therefore, in Eq.~\eq{YPyr2init_mir} all summands except the first one cancel, and that equation takes the compact form
    \be
        \log {Y_{\figp_2}^+ Y_{\figp_2}^-}
        =
        \log\frac {(1+Y_{\figf})(1+Y_{\fo_2})(1+Y_{\fO_2}) } {(1+1/Y_{\figp_3})(1+1/Y_{\figF})}.
    \ee
    Equations for $Y_{{\fo}_{n}}$, $Y_{{\fO}_{n}}$ are obtained from~\eq{Yomir},~\eq{YOmir} in a similar way with the use of new identities \eq{newlap}, and they are precisely equations \eq{YoOeq1}--\eq{YoOeq3} which were given in the introduction:
    \beqa
        \log Y_{{\fO}_{n}}^+Y_{{\fo}_{n}}^-
        &=& \log\frac {1+Y_{\figp_n}}{(1+1/Y_{\fo_{n+1}})(1+1/Y_{\fO_{n-1}})} \,\, , \,\, n>1\\
        \log {Y_{{\fo}_{n}}^+Y_{{\fO}_{n}}^-}
        &=& \log\frac {1+Y_{\figp_n}}{(1+1/Y_{\fO_{n+1}})(1+1/Y_{\fo_{n-1}})} \,\, , \,\, n>1,
    \eeqa
    while for $n=1$
    \beq
        \log Y_{{\fO}_{1}}^+Y_{{\fo}_{1}}^-
        = \log\frac {1+Y_{\fF}}{(1+1/Y_{\fo_{2}})} \,\, ,\;\;
        \log {Y_{{\fo}_{1}}^+Y_{{\fO}_{1}}^-}
        = \log\frac {1+Y_{\fF}}{(1+1/Y_{\fO_{2}})} \,\,.
    \eeq

\subsection{Integral equations for excited states}
As we have shown above the equations \eq{YFmir}-\eq{YOmir}
contain important structural information about the Y-system.
However, those equations do not make much sense
when understood literally since they describe the ground
state which is protected by super-symmetry, and the
Y-functions are degenerate in this case. There is a way to extend
these equations to excited states, with the Y-functions becoming very nontrivial.
For the case $Y_{{\fo}_a} = Y_{{\fO}_a} = Y_{{\fm}_a}$, $u_{4,j} = u_{\bar 4,j}$
similarly to \cite{Gromov:2009bc} we propose, as a conjecture, the following set of equations:
\beqa
\label{YFE}
    \log Y_{\figF} &=&
    +K_{m-1}*\log\frac{1+1/Y_{\figb_{m}}}{1+Y_{\figp_{m}}}
    +2\mathcal{R}_{1m}^{(01)}*\log(1+Y_{\figm_m})
    +
    2\left[\log\frac{R_4^{(+)}}{{R_4^{(-)}}}\right]
    + i\pi
\\
\label{YfE}
    \log Y_{\figf }&=&
    -K_{m-1}*\log\frac{1+1/Y_{\figb_{m}}}{1+Y_{\figp_{m}}}
    -2\mathcal{B}_{1m}^{(01)}*\log(1+Y_{\figm_m})
    -
    2\left[\log\frac{B_4^{(+)}}{{B_4^{(-)}}}\right]
    -i\pi
\\
\label{YpE}
    \log Y_{{\figp}_{n}}&=&-K_{n-1,m-1}*\log(1+Y_{{\figp}_{m}})
    -K_{n-1}*\log\frac{1+Y_{\figF}}{1+1/Y_{\figf}}
\\
\nn
    &+&
    2\left({\cal R}^{(01)}_{nm}+{\cal
    B}^{(01)}_{n-2,m}\right)*\log(1+Y_{\figm_m})
\\
\nn
    &+&
    2\left[\sum\limits_{k=-\frac{n-1}{2}}^{\frac{n-1}{2}}
    \log\frac{R_4^{(+)} (u+ik)}{{R_4^{(-)} (u+ik)}} \right]
    +
    2\left[\sum\limits_{k=-\frac{n-3}{2}}^{\frac{n-3}{2}}
    \log\frac{B_4^{(+)} (u+ik)}{{B_4^{(-)} (u+ik)}} \right]
\\
\label{YbE}
    \log Y_{{\figb}_{n}}&=&K_{n-1,m-1}*\log(1+1/Y_{{\figb}_{m}})
    +K_{n-1}*\log\frac{1+Y_{\figF}}{1+1/Y_{\figf}}
\\
\label{YmE}
    \log Y_{{\figm}_{n}}&=&
    J \log\frac{x^{[-n]}}{x^{[+n]}}
    -
    {\cal B}_{n1}^{(10)}*\log(1+1/Y_{\figf})
    +{\cal R}_{n1}^{(10)}*\log(1+Y_{\figF})
\\
\nn
    &+&\left({\cal R}_{nm}^{(10)}+{\cal B}_{n,m-2}^{(10)}\right) * \log(1+Y_{\figp_m})
\\
\nn
    &+&
    \(2\tilde{\cal S}_{nm} - {\cal R}_{nm}^{(11)} + {\cal B}_{nm}^{(11)}\) * \log(1+Y_{\figm_m})
    + \left[ \sum_{k=-\frac{n-1}{2}}^{\frac{n-1}{2}} \log \Phi_4(u+ik)\right]+i\pi n
\eeqa
where $J = L + K_4$, $\Phi_4$ is given by \eq{Phi4} (with $B^{\pm}_1, B^{\pm}_3$ in that expression replaced by unity)
and the exact positions of the Bethe roots are determined by
\be \la{EBAE}
    Y^{\ph}_{\figm_1} (u_{4,j}) = -1, \ \ j = 1,\dots,K_4\;.
\ee
The label ``$\ph$" here means that one should analytically continue
the equation for $Y_{\figm_1}$ to the physical
sheet, like it was done for the first time in \cite{Gromov:2009zb}. The Bethe roots are additionally constrained by a condition imposed on total momentum (the trace cyclicity condition). We can write this constraint in a form similar to~\eq{Eintro}:
    \beq
    \sum_{a=1}^{\infty}\int_{-\infty}^{\infty}\frac{du}{2\pi i}\frac{\d p_a^{\rm mir}(u)}{\d u}
    \log(1+Y^{\rm mir}_{\fo_a})(1+Y^{\rm mir}_{\fO_a})
    +\sum_{j=1}^{K_4}p^{\rm ph}(u_{4,j})+\sum_{j=1}^{K_{\bar4}}p^{\rm ph}(u_{\bar4,j})
    = 2\pi m,\ m \in \mathbb{Z}
    \eeq
(recall that the momentum $p(u)$ was introduced in~\eq{pepsDef}).
This expression can be simplified in our case, as $Y_{{\fo}_a} = Y_{{\fO}_a} = Y_{{\fm}_a}$ and $u_{4,j} = u_{\bar 4,j}$.

Note that in equations for excited states, in the terms without convolutions the branch $x^{\ph}$ should be used for $x(u_{4,j})$ and $x^{\mir}$ should be used for $x(u)$ with $u$ being the free variable.

Strictly speaking these equations are only valid for some
particular values of $\lambda$ and configurations of roots.
In other cases the equations may require some modification.
This question is usually subjected to case-by-case study (see e.g. \cite{Dorey:1997rb, Dorey:1996re,Bazhanov:1996aq}).

In general the procedure is the following - one can start from a
sufficiently large $L$ or small $\lambda$ where the terms with $\log(1+Y_{\fo_n})$
are irrelevant and the asymptotic solution of \cite{Gromov:2009zz} should be a good
approximation. The condition \eq{EBAE} can be discarded for a while, and one should find such a configuration of the roots $u_{4,j}$
(usually they are sufficiently
close to the origin in this case) that the asymptotic solution satisfies the equations for excited states we proposed
above. After that the equations should be analytically continued in $L,\lambda$ and $u_{4,j}$.

This procedure in general is rather complicated, however our experience with
the Konishi operator in ${\rm AdS}_5/{\rm CFT}_4$ \cite{Gromov:2009zb} tells us that one can probably use the equations
above as they are from $\lambda=0$ to very large $\lambda$'s.
At the same time the Y-system functional equations are not affected by
these modifications and they are more suitable for the strong coupling
analysis \cite{Gromov:2009tq}. Moreover, they are not restricted to the ``${\mathfrak sl}(2)$"
subsector.

The possibility that some singularities could collide with the integration contours
and modify the equations
when some parameters (such as the coupling) are changed was studied in detail in \cite{Dorey:1997rb, Dorey:1996re,Bazhanov:1996aq}. For AdS/CFT, this issue was mentioned in \cite{Gromov:2009zb}, and
following that proposal, such a possibility was explored
in \cite{Arutyunov:2009ax} for ${\rm AdS}_5/{\rm CFT}_4$.\footnote{In \cite{Arutyunov:2009ax} an attempt was also made to estimate the ``critical" values
of the 't Hooft coupling - values for which the equations for
excited states should be modified by extra terms.
The result from \cite{Arutyunov:2009ax} is $\lambda_{\rm critical}\simeq 774$. The method used in that work
is based on the asymptotic solution \cite{Gromov:2009zz} of the Y-system.
The asymptotic solution works perfectly for very small and very large values of the coupling,
however it very badly approximates the exact Y-functions for $\lambda\sim 700$.
Thus the only reasonable estimate at the moment for the critical value is $\lambda>700$, from the results of \cite{Gromov:2009zb},
where no singularity was found in numerical studies of the
TBA equations in the range $0<\lambda<700$.}


\section{Solution of the Y-system in the scaling limit}
In this section we obtain a solution of the ${\rm AdS}_4/{\rm CFT}_3$ Y-system in the strong coupling scaling limit, considering the $sl(2)$ subsector. In this case, the Y-functions which correspond to the momentum-carrying roots are equal. We show that the spectrum obtained from the Y-system is in complete agreement with the results from quasiclassical string theory.

\subsection{Y-system equations in the scaling limit}

In the scaling limit the Y-system simplifies in several important ways. In this section and section 5 we use rescaled rapidities $z = \frac{u}{2h(\lambda)}$ (similarly to \cite{Gromov:2009tq}), and since $h(\lambda)\to\infty$, we can neglect shifts in the arguments in the l.h.s. of the Y-system equations\footnote{This simplification of the Y-system involves certain subtleties, as the shifts in the argument of the Y-functions cannot be neglected close to the branch cuts. This issue can be treated in our case in the same way as in \cite{Gromov:2009tq}.}. Hence with $1/h^2$ precision the Y-system  becomes a set of algebraic, instead of functional, equations. Moreover, for the $sl(2)$ subsector $Y_{\fo_a}=Y_{\fO_a}$. Also, only $z_{4,k}$ and $z_{\bar 4,k}$ Bethe roots are introduced (see \eq{Yismin1}), and they coincide pairwise: $z_{4,k} = z_{\bar 4, k}$. Denoting $Y_{\fm_a} \equiv Y_{\fo_a}$ we get three infinite series of equations
\beqa
\la{ie1}
    Y_{\fb_s}^2&=&(1+Y_{\fb_{s+1}})(1+Y_{\fb_{s-1}})\;\;,\;\;s=3,4,\dots\;,
\\
\la{ie2}
    Y_{\fp_a}^2&=&\frac{(1+Y_{\fm_a})^2}
    {(1+1/Y_{\fp_{a+1}})(1+1/Y_{\fp_{a-1}})}\;\;,\;\;
    a=3,4,\dots\;,
\\
\la{ie3}
    Y_{\fm_a}^2&=&\frac{(1+Y_{\fp_a})}{(1+1/Y_{\fm_{a+1}})(1+1/Y_{\fm_{a-1}})}\;\;,
    \;\;a=2,3,\dots\;,
\eeqa
plus four more equations
\beqa
\la{ne1}
    Y_{\fp_2}^2&=&
    \frac{(1+Y_{\ff})(1+Y_{\fm_2})^2}{(1+1/Y_{\fp_3})(1+1/Y_{\fF})}\;,\ \
\\
\la{ne2}
    Y_{\fb_2}^2&=&\frac{(1+Y_{\fb_3})(1+Y_{\fF})}{(1+1/Y_{\ff})}\;,
\\
\la{ne3}
    Y_{\fF}^2&=&\frac{(1+Y_{\fb_2})(1+Y_{\fm_1})^2}{(1+1/Y_{\fp_2})}\;,\ \
\\
\la{ne4}
    Y_{\fm_1}^2&=&\frac{(1+Y_{\fF})}{(1+1/Y_{\fm_{2}})}\;.
\eeqa

Together with the Y-system, we have to solve the non-local equation
\beq
     \log Y_{\figF}Y_{\figf } =
    2 \sum_{m=1}^{\infty} (\mathcal{R}_{1m}^{(01)} - \mathcal{B}_{1m}^{(01)})*\log(1+Y_{\figm_m})
    +
    2\log\frac{R_4^{(+)} B_4^{(-)}} {{R_4^{(-)}} B_4^{(+)}},
\eeq
which can be obtained by adding up \eq{YfE} and \eq{YFE} (it corresponds to the gray node in Fig. 1, right). Introducing the following notation
\beq
    G_k(x)=\frac{1}{h}\sum_{j}^{M_k}
    \frac{1}{x-x_{k,j}}\frac{x_{k,j}^2}{x_{k,j}^2-1},\;\ \
    k = 4, \bar 4
\eeq
\beq
    f_k(z)=\exp\Big(-i G_{k}\big(x(z)\big)\Big)
    \;\;,\;\;
    \bar f_k(z)=\exp\Big(+i G_k\big(1/x(z)\big)\Big)\;, \la{fs}
\eeq
where the mirror branch of $x$ is used for $x(z)$ and the physical branch for $x_{k,j}$ (this choice of branches is used by default in sections 4 and 5), following \cite{Gromov:2009tq}
we can write  the non-local equation in the form
\be
\label{nonlocStr}
    F = \frac{1}{f \bar f} \prod_{n=1}^{\infty} (1+Y_{\fm_n})^2
\ee
where
\be
\label{deffz}
    F \equiv Y_{\ff} Y_{\fF},\ \ f(z) = f^2_4(z), \ \ \bar f(z) = \bar f^2_4(z).
\ee

Note that, similarly to \cite{Gromov:2009tq}, the Bethe roots have to satisfy the constraint
\be
    \cQ_1 =2\pi m+{\cal O}(1/h), \ \ m\in{\mathbb Z}\;,
\ee
where
\beqa\la{zeromom0}
    \cQ_1 &=&
    \sum_{j=1}^{M_4}         \frac{1}{i}\log\frac{x_{4,j}^+}{x_{4,j}^-} +
    \sum_{j=1}^{M_{\bar4}}\frac{1}{i}\log\frac{x_{\bar 4,j}^+}{x_{\bar 4,j}^-}
   \simeq
    \sum_{j=1}^{M_4}       \frac{x_{4,j}}{h(x_{4,j}^2-1)}+
    \sum_{j=1}^{M_{\bar 4}}\frac{x_{\bar 4,j}}{h(x_{\bar 4,j}^2-1)}.
\eeqa
This condition reflects the cyclicity symmetry of single trace operators. Its consistency with the other equations remains to be checked, and we assume that equation to be satisfied.

\subsection{Asymptotics of Y-functions}
The Y-system equations should be supplemented by boundary conditions on the functions $Y_{a,s}$, i.e. by their large $a,s$ asymptotics. In our case, similarly to ${\rm AdS}_5/{\rm CFT}_4$ (see \cite{Gromov:2009tq}), we demand that $\Yp{a}$ and $\Ym{a}$ have the same asymptotics as the $L\to\infty$ solution, which was constructed in section 2.  As for the functions $\Yb{s}$, we demand that their large $s$ asymptotics is polynomial in $s$, which is true for the $L\to\infty$ solution as well.


%

It is straightforward to show that the expressions for Y-functions from section 2 can be recast in the following form:
\be
    {\bf Y}_{\figm_a} = (-1)^a \Delta^a
    \frac{f(\bar f-1)^2 \bar f^{a} - \bar f(f-1)^2 f^{a}}
    {f \bar f(\bar f - f)}
\ee
\beq
\la{Ynw}
    {\bf Y}_{\fb_s}(z)=\(s-A\)^2-1\;\;,\;\;
    {\bf Y}_{\fp_a}
    =\frac{(T-1)^2 S T^{a-1}}{(S T^{a+1}-1)(S T^{a-1}-1)}\;,
\eeq
where
\beq
\la{A0}
    A=\frac{1}{\bar{f}-1}+\frac{f}{f-1}\;\;,\;\;
    S=\frac{\bar f(f-1)^2}{f(\bar f-1)^2}\;\;,\;\;
    T=\frac{f}{\bar f}\;,
\eeq
\be
\label{defdelta}
    \Delta=\exp\(-i\frac{L x/h-{\cal Q}_1}{x^2-1}\)
\ee
while $f, \bar f$ are given by \eq{deffz}. As for real $z$ the quantity ${f(z)}/{\bar f(z)}$ is a pure phase, to investigate the large $a,s$ limit we consider Y-functions of shifted argument $z-i0$. We have then
$\left| {f(z-i0)}/{\bar f(z-i0)}\right| > 1$ and we get
\beq\la{bp}
\lim_{a\to\infty}\frac{\log {\bf Y}_{\fp_a}(z-i0)}{a}=\log \frac{\bar f}{f}\;.
\eeq
Similarly,
\beq\la{bm}
\lim_{a\to\infty}\frac{\log {\bf Y}_{\fm_a}(z-i0)}{a}=\log \(-\Delta f\)\;.
\eeq
Conditions~\eq{bp},~\eq{bm} are the boundary conditions which we impose on the functions $\Yp{a}, \Ym{a}$ for finite $L$ at strong coupling.

\subsection{Solution in upper and right wings}
In this section, we solve the Y-system partially, expressing the upper wing functions $\Yp{a}$, $\Ym{a}$ and the right wing functions $\Yb{s}$ in terms of only three yet unknown functions.

Using the analogy between our Y-system and the one considered in \cite{Gromov:2009tq}, it can be shown that the functions $\Yp{a}, \Ym{a}$ can be constructed in the following way:
\be
\label{TtoYsl2vert}
    1 + \Yp{a} = \frac{T^2_{a,0}}{T_{a+1,0} T_{a-1,0}}, \
    1 + \Ym{a} = \frac{T^2_{a,1}}{T_{a+1,1} T_{a-1,1}},
\ee
where the set of functions $T_{a,s}$, which is the general solution of the Hirota equation in the vertical strip, was found in \cite{Gromov:2009tq}.
Those functions are:
\beqa
    T_{a,2} &=& 1
\\
    T_{a,1} &=&
    \frac{y_1y_2}{(y_1-y_2)(y_1y_2-1)}
    \(\frac{y_1}{y_1^2-1}\(S_1y_1^{a}+\frac{1}{S_1y_1^{a}}\)
    -\frac{y_2}{y_2^2-1}\(S_2y_2^{a}+\frac{1}{S_2y_2^{a}}\)\)
\\
    T_{a,0} &=& (T^2_{a,1} - T_{a+1,1}T_{a-1,1})/T_{a,2}.
\eeqa
with $y_1, y_2, S_1$ and $S_2$ being arbitrary parameters.
The functions $\Ym{a},\Yp{a}$, given by \eq{TtoYsl2vert},
satisfy the Y-system equations~\eq{ie2} and \eq{ie3} for arbitrary $y_1(z), y_2(z), S_1(z)$ and $S_2(z)$. The asymptotic conditions~\eq{bp},\eq{bm} fix $y_1$ and $y_2$:
\beq
    y_1=-\frac{f_4}{\bar f_4}\;\;,\;\;y_2=\Delta f_4 \bar f_4\;\;.
\eeq

The general solution of \eq{ie1} with polynomial large $s$ asymptotics is (see \cite{Gromov:2009tq})
\be
\label{ybViaAz}
    Y_{\fb_s}(z)=\(s-A(z)\)^2-1\;\;,\;\;
\ee
with arbitrary $A(z)$. Thus, the solution of our Y-system in the upper and right wings is expressed in terms of three unknown functions $S_1(z), S_2(z)$ and $A(z)$.

\subsection{Matching wings}

By now, we have constructed $\Yb{n}, \Yp{n}$ and $\Ym{n}$ for all $n$ in terms of $A(z), S_1(z)$ and $S_2(z)$. To find those three functions, as well as $\Yf$ and $\YF$, we have to solve the five remaining equations~\eq{ne1},\eq{ne2},\eq{ne3},\eq{ne4},\eq{nonlocStr}. Excluding $\Yf$ and $\YF$, we get:
\beqa
\la{we1}&&
    \frac{Y_{\fp_2}^2(1+1/Y_{\fp_3})}{(1+Y_{\fm_2})^2}=\frac{F}{(A-1)^2}\;,\\
\la{we2}&&
    \frac{1+1/Y_{\fp_2}}{(1+Y_{\fm_1})^2}=\frac{1}{A^2}\(\frac{(A-1)^2}{F}-1\)^2\;,\\
\la{we3}&&
    Y_{\fm_1}^2(1+1/Y_{\fm_2})=-\frac{(A-1)^2 (F-1)} {F - (A-1)^2}  \;,
\eeqa
\be
    \frac{1}{f^2_4 \bar f^2_4} \prod_{n=1}^{\infty} (1+Y_{\fm_n})^2 = F
\ee

The r.h.s. of the four equations above depends only on $F(z)$ and $A(z)$, and they can be solved perturbatively in $\Delta$, like analogous equations in~\cite{Gromov:2009tq}. Namely, we find several terms in the expansion of unknown functions in powers of $\Delta$, notice a simple relation between consecutive terms, and sum up the series in $\Delta$ assuming this relation to hold for all terms\footnote{For $\Delta=0$, there are several solutions of Eqs.~\eq{we1}-\eq{we3}. We choose the one consistent with the asymptotic solution of Y-system.}. It is then easy to check that the functions obtained in this way are indeed solutions of~\eq{we1}, \eq{we2},~\eq{we3}. The result is:
\beqa
\nn
    A &=& \frac{(1 + \Delta) ((1 - f_4 \bar f_4)(1-\Delta^2f_4 \bar f_4)-
    \Delta (f_4 - \bar f_4)^2)
    ((1 + f_4 \bar f_4)(1+\Delta^2f_4 \bar f_4)-
    \Delta (f_4 + \bar f_4)^2)}
     {(-1 + \Delta) (f-1) (\bar f-1)
     (\Delta^2 f - 1) (\Delta^2 \bar f -1)}
\\
\\
    S_1 &=& \frac{(f - 1)\bar f
    (\Delta^2 \bar f-1)}
    {f (\Delta^2 f-1)(\bar f-1)}
\\
    S_2 &=& \frac{(f-1)(\bar f-1)}
    {f \bar f(\Delta^2 f-1) (\Delta^2 \bar f-1)}
\\
    F &=&
    \frac{(\Delta f-1)^2 (\Delta \bar f-1)^2}
    {(\Delta-1)^4 f \bar f}
\eeqa
Putting those functions into the expressions for the Y-functions \eq{TtoYsl2vert}, \eq{ybViaAz}, we obtain all the $Y_{a,s}$ in terms of $f, \bar f$ and $\Delta$ (using~\eq{ne4} to find $\YF$ and then getting $\Yf$ from $F=\Yf\YF$).
\newline

Recalling the definitions \eq{deffz}, \eq{defdelta} of $f, \bar f$ and $\Delta$, we see that we have found all the Y-functions in terms of the Bethe roots. We present our solution of the Y-system in \textit{Mathematica} form in Appendix C.

    \subsection{The spectrum from Y-system}
    Here we repeat the arguments
    of \cite{Gromov:2009tq} to find the equation for the displacement
    of Bethe roots due to the finite size effects at strong coupling.
    In this section we assume $Y_{\fO_n}=Y_{\fo_n}\equiv Y_{\fm_n}$ (the situation
    where this is not the case is considered in the next section).
    We again start from the TBA equation for the momentum-carrying node
\beq\la{eqmid}
\log Y_{\fm_1}={\cal T}_{1m}*\log (1+Y_{\fm_m})+{\cal R}^{(10)}\cut\log(1+Y_{\fF})
+
{\cal R}^{(10)}\cut K_{m-1}*\log(1+Y_{\fp_m})+i\Phi\;,
\eeq
where ${\cal T}_{1m} \equiv 2\tilde{\cal S}_{nm} - {\cal R}_{nm}^{(11)} + {\cal B}_{nm}^{(11)}$ and $\Phi$ represents extra potentials in the TBA equations
for the excited states.
The only difference with \cite{Gromov:2009tq}
is absence of $2$'s in front of the second and third terms.
Thus we can use the same trick as in \cite{Gromov:2009tq}
to get the expression for $Y_{\fm_1}$ in physical kinematics:
\beqa\la{YmTBA}
\log \frac{Y_{\fm_1}^\ph}{Y_{\fm_1}^{\ph0}}&=&{\cal T}^{\ph,\mir}_{1m}*\log (1+Y_{\fm_m})+
{\cal R}^{(10)\ph,\mir}\cut\log\(\frac{1+Y_{\fF}}{1+Y_{\fF}^0}\)\\
\nn&+&
{\cal R}^{(10)\ph,\mir}\cut K_{m-1}*\log\(\frac{1+Y_{\fp_m}}{1+Y_{\fp_m}^0}\)
+K_{m-1}(z_k-\tfrac{i}{4h})*\log\(\frac{1+Y_{\fp_m}}{1+Y_{\fp_m}^0}\)\;.
\eeqa
Now we simply have to expand the kernels at large $h$ and
substitute Y's. Let us denote
\beqa
\nn r(x,z)&=&\frac{x^2}{x^2-1}\frac{\d_z}{2\pi h}\frac{1}{x-x(z)}\;\;,\;\;
u(x,z)=\frac{x}{x^2-1}\frac{\d_z}{2\pi h}\frac{1}{x^2(z)-1}
.
\eeqa
We rearrange
the terms in \eq{YmTBA} to
evaluate the following ``magic" products\footnote{To compute
these products we again use the $z-i0$ prescription to ensure their convergence.
This prescription is inherited from the TBA equation for excited states where the integration
should go slightly below the real axis.}
\beqa
\nn e^{+{\cal M}_0}&\equiv&\prod_{m=1}^\infty (1+Y_{\fm_m})^{2m}=
    \frac{ (\Delta^2 f-1 )^4  (\Delta^2 \bar f-1 )^4}
    {(\Delta-1)^6 (\Delta+1)^2 (\Delta f+1)^2 (\Delta \bar f+1)^2  (\Delta^2 f
       \bar f-1 )^2}
\\
\nn e^{-{\cal M}_+}&\equiv&\frac{1+Y_{\fF}}{1+Y_{\fF}^0}\prod_{m=2}^\infty\(\frac{1+Y_{\fp_m}}{1+Y_{\fp_m}^0}\)^m\prod_{m=1}^\infty \frac{1}{(1+Y_{\fm_m})^{2m}}
=-\frac{(f\Delta+1)^2(f\bar f\Delta^2-1)}{(f\Delta^2-1)^2}
\\
\la{Mm} e^{+{\cal M}_-}&\equiv&\frac{1+1/Y_{\ff}^0}{1+1/Y_{\ff}}\prod_{m=2}^\infty\(\frac{1+Y_{\fp_m}}{1+Y_{\fp_m}^0}\)^{m-2}
=-\frac{(\bar f\Delta^2-1)^2}{(\bar f\Delta+1)^2(f\bar f\Delta^2-1)}.
\eeqa
We get the following corrected Bethe equation for the ${\mathfrak sl}_2$
sector
\beqa\la{BAEcor}
\nn1&=&\(\frac{x_k^-}{x_k^+}\)^L\prod_{j=1}^M\frac{x_k^--x_j^+}{x_k^+-x_j^-}\frac{1-1/(x_k^+x_j^-)}{1-1/(x_k^- x_j^+)}
\sigma^2(z_k,z_j)\\
&\times&\exp\[-\int_{-1}^1 \Big(r(x_k,z){\cal M}_+
-r(1/x_k,z){\cal M}_-
+u(x_k,z){\cal M}_0\Big)dz\]\;,
\eeqa
and the equation for the energy at strong coupling is
\beq\la{energyY}
E=\sum_{i=1}^M\frac{x_i^2+1}{x_i^2-1}+
\int_{-1}^1
\frac{dz}{4\pi}\frac{z}{\sqrt{1-z^2}}\d_z{\cal M}_0\;.
\eeq
The extra factor of $2$ in the denominator under the integral is due to the
single magnon dispersion relation,
which includes an extra $1/2$ compared to ${\rm AdS}_5/{\rm CFT}_4$.

\subsection{Non-symmetric strong coupling solution}
In this section we present a simple strong coupling
solution with $Y_{\fo}\neq Y_{\fO}$. It can be used as a test of the
new structure of the Y-system which we proposed in the introduction.

We consider the limit where the massive nodes are completely decoupled from the
rest of the system. We solve the following infinite set of equations
\beqa
     Y_{{\fO}_{n}}Y_{{\fo}_{n}}
    &=& \frac {1}{(1+1/Y_{\fo_{n+1}})(1+1/Y_{\fO_{n-1}})} \,\, , \,\, n>1\\
     {Y_{{\fo}_{n}}Y_{{\fO}_{n}}}
    &=& \frac {1}{(1+1/Y_{\fO_{n+1}})(1+1/Y_{\fo_{n-1}})} \,\, , \,\, n>1.
\eeqa
The explicit general solution of this system with two parameters $\alpha$ and $\beta$
is
\beq
Y_{\fo_n}=
\left\{\bea{ll}
\frac{\left(\alpha ^2-1\right)^2 \alpha ^n}{\beta  \left(\alpha ^n
   -\frac{\alpha +\beta}{\alpha\beta+1}\right)\left(\alpha +\frac{1}{\beta }\right) \left(\alpha ^{n+2} -
   \frac{\alpha\beta  +1}{\alpha +\beta }\right)(\alpha +\beta )}&\;,\;\;n \ \text{is odd}\\
\frac{\left(\alpha ^2-1\right)^2 \alpha ^n}{\left(\alpha ^n-1\right)\left(\alpha +\frac{1}{\beta }\right)
   \left(\alpha ^{n+2}-1\right)
   (\alpha +\beta )}&\;,\;\;n\ \text{is even}\\
\eea\right.
\eeq
and $Y_{\fO_n}$ is obtained from $Y_{\fo_n}$ by replacing $\beta\to 1/\beta$. We can easily compute ${\cal M}_0$
for this solution,
\beq
e^{{\cal M}_0}=\frac{(\alpha\beta+1)(\alpha+\beta)}{\left(\alpha
   ^2-1\right)^2 \beta }
\eeq
and by matching with the asymptotic solution we identify
\beq
\lambda_1=\alpha\;\;,\;\;\lambda_2=\lambda_3=\lambda_4=0\;\;,\;\;\lambda_5=-\beta\;.
\eeq
so that we get
\beq\la{neqY}
e^{{\cal M}_0}=\frac{(\lambda_5-\lambda_1)(1-\lambda_1\lambda_5)}{\left(\lambda_1^2-1\right)^2 \lambda_5 }\;.
\eeq

    \section{One-loop strong coupling quasi-classical string spectrum}
    \FIGURE[ht]{\la{figst}
    \includegraphics[scale=0.8]{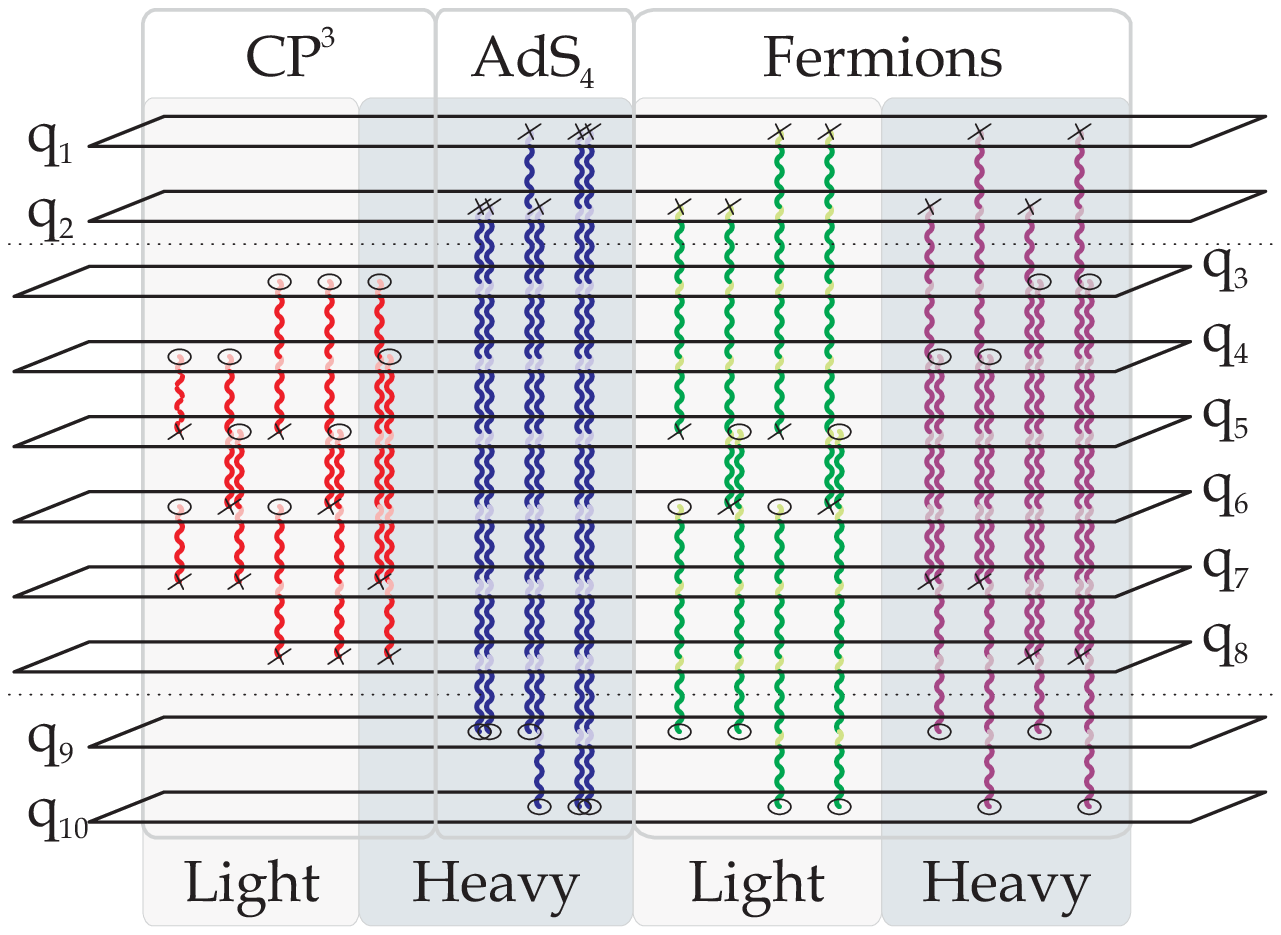}
    \caption{Elementary excitations of the string in ${\rm AdS}_4\times {\mathbb{CP}}^3$}
    }
    In this section we briefly describe the construction of
    \cite{Gromov:2009tq}, applied for the ABJM model. The algebraic curve
    described in \cite{GV1} can be used to compute the one-loop correction
    for a generic finite gap classical string state by computing the spectrum of fluctuations around
    a given solution.
    We assume that the one-loop
    shift computed from the algebraic curve agrees with the strong coupling expansion
    of ABA in the limit $L/h\gg 1$. This assumption was explicitly verified for the folded string
    in \cite{Gromov:2008fy}. The general proof like in \cite{Gromov:2007ky} is still missing.

    There is yet another way to compute the one-loop shift directly from the
    world-sheet action which is similar to the algebraic curve computation.
    Whereas for the folded string both computations give the same excitation frequencies\footnote{similar analysys for the giant magnon was done in \cite{Sh, Bombardelli:2008qd}} \cite{McLoughlin:2008ms,Gromov:2008fy},
    for the circular string a negative result was obtained in \cite{McLoughlin:2008he}. Recently
    it was shown that one should be more carefull with the periodicity of the fermionic fields
    in the world-sheet approach and the corrected derivation leads to agreement with the algebraic
    curve frequencies \cite{Mikhaylov:2010ib} so we assume all approaches to be consistent with each other.

    The pattern of excitations in the ABJM theory is quite different from
    that of ${\rm AdS}_5\times {\rm S}^5$.
    The string in ${\rm AdS}_4\times {\mathbb{CP}}^3$ has $8$ bosonic ($3$ modes of ${\rm AdS}_4$ and $5$ of ${\mathbb{CP}}^3$) and
    $8$ fermionic excitations. They are divided into heavy and light modes (see Fig.\ref{figst}).
    The dispersion relations for the heavy and light modes differ by a factor of two. We will see that this complicated structure of
    heavy and light fluctuations is captured by the Y-system.
    As usual the one-loop shift is given by a sum over the fluctuation energies \cite{Frolov:2002av}.
    In the algebraic curve language the fluctuations are the small cuts (i.e. poles)
    connecting different sheets of the algebraic curve $q_i$. The poles could be placed only in certain special
    positions $x_n^{ij}$ given by
    \beq
    q_i(x_n^{ij})-q_j(x_n^{ij})=2\pi n\;.
    \eeq
    The quasiclassical Bohr-Sommerfeld quantization condition constrains the minimal residue of the pole.
    Insertion of the pole results in displacement of the other singularities.
    Moreover the pole by itself carries the energy $\omega(x_n^{ij})$ for the light mode and
    $2\omega(x_n^{ij})$ for the heavy mode where
    \beq
    \omega(x)=\frac{1}{x^2-1}.
    \eeq
    Following \cite{Gromov:2009tq} we first compute this second part of the one-loop shift
    which does not take into account the back-reaction of the fluctuation
    on the large cuts. Then we have
    \beq\la{sumE}
    \delta E_{1-loop}^0=\frac{1}{2}\sum_n\sum_{\rm light}(-1)^{F_{ij}}\omega(x_n^{ij})+
    \sum_n\sum_{\rm heavy}(-1)^{F_{ij}}\omega(x_n^{ij})
    \eeq
    where $F_{ij}=1$ for bosonic modes and $-1$ for fermionic. The modes are
    \beqa
    (i,j)&=&(4,5),(4,6),(3,5),(3,6)\;\;,\;\;{\rm light\;bosonic\;modes}\\
    (i,j)&=&(3,7),(2,9),(1,9),(1,10)\;\;,\;\;{\rm heavy\;bosonic\;modes}\\
    (i,j)&=&(2,5),(2,6),(1,5),(1,6)\;\;,\;\;{\rm light\;fermionic\;modes}\\
    (i,j)&=&(2,7),(1,7),(2,8),(1,8)\;\;,\;\;{\rm heavy\;fermionic\;modes}
    \eeqa
    Rewriting the sum \eq{sumE} as an integral we get
    \beq\la{Eint}
    \delta E_{1-loop}^0=\sum_{(ij)}\cw\oint_{{\mathbb U}^+}\frac{dx}{2\pi i}\omega(x)\d_x{\cal N}_0
    \eeq
    where the integration goes over the upper half of the unit circle $|x|=1$ and we denote
    \beq
    \label{eN0prod}
    e^{{\cal N}_0}=\prod_{\rm light}(1-e^{-i p_i+i p_j})^{F_{ij}}\prod_{\rm heavy}(1-e^{-i p_i+i p_j})^{2F_{ij}}
    \eeq
    Note that $\d_x{\cal N}_0$ is constructed to have the residue $\pm 1$ ($\pm2$) exactly at the position of the light (heavy) mode
    $x_n^{ij}$.
    More explicitly we can write
    \beq\la{N0}
    e^{{\cal N}_0}=\frac{ (\lambda_1^2-1 )^2(\lambda_2^2-1 )^2 (\lambda_1 \lambda_2-1)^2   (\lambda_3
       \lambda_4-1)^2 (\lambda_3-\lambda_5) (\lambda_4-\lambda_5) (\lambda_3 \lambda_5-1) (\lambda_4 \lambda_5-1)}{(\lambda_1
       \lambda_3-1)^2 (\lambda_2 \lambda_3-1)^2 (\lambda_1 \lambda_4-1)^2 (\lambda_2 \lambda_4-1)^2 (\lambda_1-\lambda_5)
       (\lambda_2-\lambda_5) (\lambda_1 \lambda_5-1) (\lambda_2 \lambda_5-1)}
    \eeq
    where
    \beq
    \lambda_a=e^{-i q_a}\;.
    \eeq
    For $sl_2$ sector \cite{GV2} there are only cuts connecting 2nd and 9th sheets
    so that
    \beq\la{sl2}
    \lambda_5=1\;\;,\;\;\lambda_3=\lambda_4=\Delta
    \;\;,\;\;\lambda_2=\Delta f\;\;,\;\;\lambda_2=\Delta \bar f
    \eeq
    and \eq{N0} simplifies to
    \beq
    e^{{\cal N}_0}=
    \frac{(\Delta-1)^6 (\Delta+1)^2 (\Delta f+1)^2 (\Delta \bar f+1)^2  (\Delta^2 f
       \bar f-1 )^2}{ (\Delta^2 f-1 )^4  (\Delta^2 \bar f-1 )^4}\;,
    \eeq
    where we recognize ${\cal N}_0=-{\cal M}_0$ and \eq{Eint} coincides precisely with the
    second term in \eq{energyY}!

    Then one should also take into account the back-reaction of the fluctuations.
    As it is explained in detail in \cite{Gromov:2009tq} for that one should work with
    the modified Bethe equations.
 	We need to consider only the fluctuations touching those sheets where the macroscopic
    cuts are located. Hence, one of those sheets has to be the 2nd or the 9th sheet.
    Computing the r.h.s of~\eq{eN0prod} with that restriction imposed, we get, similarly to \cite{Gromov:2009tq},
    \beqa
    e^{{\cal N}_+}&=&-\frac{ (\lambda_1 \lambda_2-1 )  (\lambda_2^2-1 ){}^2 }{ (\lambda_2 \lambda_3-1 )
        (\lambda_2 \lambda_4-1 )  (\lambda_2-\lambda_5 )  (\lambda_2 -1/\lambda_5 )}\\
    e^{{\cal N}_-}&=&-\frac{ (\lambda_1^2-1 ){}^2  (\lambda_1 \lambda_2-1 ) }{ (\lambda_1 \lambda_3-1 )
        (\lambda_1 \lambda_4-1 )  (\lambda_1-\lambda_5 )  (\lambda_1 -1/\lambda_5 )}
    \eeqa
    and from \eq{sl2} we get
    \beq
    e^{{\cal N}_+}=-\frac{(\Delta f+1)^2  (\Delta^2 f \bar f-1 )}{ (\Delta^2 f-1 )^2}\;\;,\;\;
    e^{{\cal N}_-}=-\frac{(\Delta \bar f+1)^2  (\Delta^2 f \bar f-1 )}{ (\Delta^2 \bar f-1 )^2}
    \eeq
    and we recognize exactly the same structures \eq{Mm} we got from solving the Y-system!

    Finally by putting $\lambda_2=\lambda_3=\lambda_4=0$ we obtain
        \beq
    e^{{\cal N}_0}=\frac{ (\lambda_1^2-1 )^2 \lambda_5  }{(\lambda_1-\lambda_5)
       (\lambda_1 \lambda_5-1)}
    \eeq
which is again precisely the quantity $e^{-{\cal M}_0}$ obtained for this sector in \eq{neqY}!

We see that the nontrivial pattern of the fluctuations is reflected
in the Y-system thus providing a direct link with the worldsheet theory.
This is also a deep test of the structure of the Y-system equations we proposed.
\section{Conclusion}
In this paper we refined the Y-system for the ABJM theory which was conjectured in \cite{Gromov:2009zz}. We derived it directly through the TBA
approach and then made several highly nontrivial tests at strong coupling.
In particular we constructed the general $\mathfrak{sl}_2$
solution for the new Y-system in the scaling limit,
and also made a test for a subsector
where the difference between the old and the new Y-systems is crucial.

We also constructed the general asymptotic solution
of the Y-system for arbitrary excited states. It can be used, in particular,
for the weak coupling tests of the conjecture and as an initial configuration
for numerical iterative solutions.

{\bf Note added}
While we were working on the strong coupling solution,
the paper \cite{Bombardelli:2009xz} appeared, with a similar Y-system
and vacuum TBA equations.

\section*{Acknowledgments}

The work of NG was partially supported by the German Science Foundation (DFG) under
the Collaborative Research Center (SFB) 676 and RFFI project grant 06-02-16786. The work of F. L.-M. was partially supported by the
Dynasty Foundation (Russia) and by the grant NS-5525.2010.2.
We thank V.Kazakov, P.Vieira, V.Schomerus and Z.Tsuboi for discussions.
NG is grateful to the
Simons Center for Geometry And Physics,
where a part of the work was done, for the kind hospitality. F. L.-M. thanks DESY, where a part of this work was done, for hospitality during the 2009 summer school.

\appendix

    \section{Notation and kernels}
    We use the following notation for kernels in TBA equations:
    \beqa
        K_n (u, v) &\equiv& \frac{1}{2\pi i} \frac{\d}{\d v} \ln\frac{u-v+in/2}{u-v-in/2},
    \\
        K_{n,m}(u,v) &\equiv&
        \sum_{j=-\frac{m-1}{2}}^{\frac{m-1}{2}}\sum_{k=-\frac{n-1}{2}}^{\frac{n-1}{2}}
        K_{2j+2k+2}(u,v),
    \\
        {\cal S}_{nm}(u,v) &\equiv&
        \frac{1}{2\pi i}\frac{\d}{\d v}
        \log \sigma_{BES}(x^{[+n]}(u), x^{[-n]}(u), x^{[+m]}(v), x^{[-m]}(v))
    \\
        \tilde {\cal S}_{nm}(u,v) &\equiv& {\cal S}_{nm}(u,v) +\frac{ni}{2}{\cal P}^{(m)}(v)
    \\
        \mathcal{ B}^{(ab)}_{nm}(u,v) &\equiv&
        \sum_{j=-\frac{n-1}{2}}^{\frac{n-1}{2}}\sum_{k=-\frac{m-1}{2}}^{\frac{m-1}{2}}
        \frac{1}{2\pi i}\frac{\d}{\d v} \log
        \frac{b(u+ia/2+ij,v-ib/2+ik)}{b(u-ia/2+ij,v+ib/2+ik)}
    \\
        \mathcal{ R}^{(ab)}_{nm}(u,v) &\equiv&
        \sum_{j=-\frac{n-1}{2}}^{\frac{n-1}{2}}\sum_{k=-\frac{m-1}{2}}^{\frac{m-1}{2}}
        \frac{1}{2\pi i}\frac{\d}{\d v} \log
        \frac{r(u+ia/2+ij,v-ib/2+ik)}{r(u-ia/2+ij,v+ib/2+ik)},
    \eeqa
where
    \be
        b (u,v) = \frac{1/x^{\mir}(u) - x^{\mir}(v)}{\sqrt{x^{\mir}(v)}}, \ \
        r (u,v) = \frac{x^{\mir}(u) - x^{\mir}(v)}{\sqrt{x^{\mir}(v)}},
    \ee
    and
    \beq
    {\cal P}^{(a)}(v)=-\frac{1}{2\pi}\d_v\log\frac{x^{\mir}(v+ia/2)}{x^{\mir}(v-ia/2)}\;.
    \eeq

To derive~\eq{SBintrep} we use the following integral representation \cite{Gromov:2009bc,Arutyunov:2009kf}:
\beqa
    &&2{\cal \tilde S}_{nm}(u,v)-{\cal R}_{nm}^{(11)}(u,v)+{\cal B}_{nm}^{(11)}(u,v)
    =
    -{K}_{n,m}(u-v)\\
    &&-2\sum_{a=1}^{\infty}\int
    \left[{\cal B}^{(10)}_{n1}\left(u,w+ia/2\right){\cal B}_{1m}^{(01)}\left(w-ia/2,v\right)
     + {\cal B}^{(10)}_{n1}\left(u,w-ia/2\right){\cal B}_{1m}^{(01)}\left(w+ia/2,v\right)\right]dw \nn.
\eeqa
Equation~\eq{SBintrep} follows due to the identity ${\cal R}_{nm}^{(11)}(u,v) + {\cal B}_{nm}^{(11)}(u,v) = { K}_{n,m}(u-v)$.

\section{Fermionic duality transformation and $\mathfrak{su}(2)$}
We can transform a set of Bethe equations
into an equivalent one by application of the fermionic duality.
This follows \cite{Beisert:2005fw} closely. We construct the polynomial
\beqa\nn
P(x)&=& \prod_{j=1}^{K_4}(x-x_{4,j}^+)\prod_{j=1}^{K_{\bar 4}}(x-x_{\bar 4,j}^+)
\prod_{j=1}^{K_2}(x-x_{2,j}^-)(x-1/x_{2,j}^-)\\
&-&
\prod_{j=1}^{K_4}(x-x_{4,j}^-)\prod_{j=1}^{K_{\bar 4}}(x-x_{\bar 4,j}^-)
\prod_{j=1}^{K_2}(x-x_{2,j}^+)(x-1/x_{2,j}^+)
\eeqa
from the Bethe equations of~\cite{GV2} (given in section 2.1 of the present work) for the fermionic roots $u_1$ and $u_3$.
We see that this polynomial
has zeros for $x=1/x_{1,j}$ and $x=x_{3,j}$. Denoting the remaining
zeros of this polynomial by $\tilde x$, we get
\beq
P(x)= C \prod_{j=1}^{K_3}(x-x_{3,j})\prod_{j=1}^{K_1}(x-1/x_{1,j})
\prod_{j=1}^{\tilde K_3}(x-\tilde x_{3,j})
\prod_{j=1}^{\tilde K_1}(x-1/\tilde x_{1,j})
\eeq
(where $C$ is a constant) or in our notation (with $R\equiv R_4 R_{\bar4}, \ B\equiv B_4 B_{\bar4}$)
\beq
P(x)= C R_3B_1R_{\tilde 3}B_{\tilde 1}=\[R^{(-)} Q_2^+-R^{(+)} Q_2^-\]
\(\frac{x}{h}\)^{K_2}\prod_{j=1}^{K_4}\sqrt{x_{4,j}^+}
\eeq
\beq
\frac{P(x^+)}{P(x^-)}=\frac{R^+_3B^+_1R^+_{\tilde 3}B^+_{\tilde
1}}{R^-_3B^-_1R^-_{\tilde 3}B^-_{\tilde 1}}
= \(\frac{x^+}{x^-}\)^{K_2}
\frac{R^{(-)+} Q_2^{++}-R^{(+)+} Q_2}{R^{(-)-} Q_2-R^{(+)-} Q_2^{--}}
\eeq
and
\beq
\frac{P(1/x^-)}{P(1/x^+)}=\frac{B^-_3R^-_1B^-_{\tilde 3}R^-_{\tilde
1}}{B^+_3R^+_1B^+_{\tilde 3}R^+_{\tilde 1}}
= \(\frac{x^+}{x^-}\)^{K_2}
\frac{B^{(-)-} Q_2-B^{(+)-} Q_2^{--}}{B^{(-)+} Q_2^{++}-B^{(+)+} Q_2}
\eeq
then
\beq
f(u)=-\(\frac{x^+}{x^-}\)^{K_2}\frac{R^{(+)+}B_1^-\tilde B_1^-
B_3^+\tilde B_3^+}{R^{(-)-}B_1^+\tilde B_1^+B_3^- \tilde B_3^-}
\;\;,\;\;f_a(u)\equiv \prod_{n=-\frac{a-1}{2}}^{\frac{a-1}{2}} f(u+in)
\eeq
\beqa
T_{a,1}(u|\{u_{1,j}\},\{u_{3,j}\})&=&
f_a(u)\overline{T_{1,a}(u|\{\tilde u_{1,j}\},\{\tilde u_{3,j}\})}\\
T_{1,s}(u|\{u_{1,j}\},\{u_{3,j}\})&=&
f_s(u)\overline{T_{s,1}(u|\{\tilde u_{1,j}\},\{\tilde u_{3,j}\})}
\eeqa
Here, the bar means complex conjugation in the physical sense, i.e. the
replacement
\beq
R^{(\pm)\pm}\to R^{(\mp)\mp}\;\;,\;\;B^{(\pm)\pm}\to B^{(\mp)\mp}.
\eeq
Notice that $x$ is not inverting under this conjugation.
   \beqa
   &&{\bf Y}_{\fo_a}\simeq
\(\frac{x^{[-a]}}{x^{[+a]}}\)^{L-K_2}\overline{{\bf
T}_{a,1}}\prod_{n=-\frac{a-1}{2}}^{\frac{a-1}{2}}
   \tilde\Phi_4^{\theta^{\rm E}_{na}}(u+in)\tilde\Phi_{\bar
4}^{\theta^{\rm O}_{na}}(u+in),\\
   &&{\bf Y}_{\fO_a}\simeq
\(\frac{x^{[-a]}}{x^{[+a]}}\)^{L-K_2}\overline{{\bf T}_{a,1}}
   \prod_{n=-\frac{a-1}{2}}^{\frac{a-1}{2}}
   \tilde\Phi_4^{\theta^{\rm O}_{na}}(u+in)\tilde\Phi_{\bar
4}^{\theta^{\rm E}_{na}}(u+in)
   \eeqa
As in section 2, the factors $\tilde\Phi_{4}(u)$ and $\tilde\Phi_{\bar 4}(u)$ are constructed in such a way that
   the ABA equations for the momentum carrying nodes are given by
   ${\bf Y}_{\fo_1}^{\ph}(u_{4,j})=-1$ and ${\bf
Y}_{\fO_1}^{\ph}(u_{\bar 4,j})=-1$. We have
   \beq
   \tilde\Phi_{4}(u)=-S_4 S_{\bar 4}\frac{B_4^{(+)+}R_{
4}^{(-)-}\tilde B_1^-\tilde B_3^+}{B_4^{(-)-}R_{ 4}^{(+)+}\tilde
B_1^+\tilde B_3^-}
   e^{i{\cal Q}_1/2}\;\;,\;\;
   \tilde\Phi_{\bar 4}(u)=-S_4 S_{\bar 4}\frac{B_{\bar
4}^{(+)+}R_{\bar 4}^{(-)-}\tilde B_1^-\tilde B_3^+}{B_{\bar
4}^{(-)-}R_{\bar 4}^{(+)+}\tilde B_1^+\tilde B_3^-}
   e^{i{\cal Q}_1/2}\;.
   \eeq
where $\tilde B_l, \tilde R_l$ are defined similarly to \eq{defR}, \eq{defB}, with $x_{l,j}$ replaced by $\tilde x_{l,j}$.

\section{Explicit expressions for Y-functions}
Here we present the solution of the Y-system in the scaling limit for the $sl(2)$ sector. This solution was obtained in section 4, and below we give its explicit form, which can be used in the \textit{Mathematica} system. We denote
$\verb"d"=\Delta,\;\verb"g"=f_4,\;\verb"gb"=\bar f_4,\;\verb"Ym[a_]"=Y_{\fm_a},\;\verb"Yp[a_]"=Y_{\fp_a},\;
\verb"Yb[s_]"=Y_{\fb_s}$, and the Y-functions are given by the following code:

\bigskip

\noindent
\verb"sb={"\\
\verb"A-> ((1+d)(1-d g^2-g gb+2 d g gb-d^2 g gb-d gb^2+d^2 g^2 gb^2)"\\
\verb"    (1-d g^2+g gb-2 d g gb+d^2 g gb-d gb^2+d^2 g^2 gb^2))/"\\
\verb"    ((-1+d)(-1+g)(1+g)(-1+d g)(1+d g)(-1+gb)(1+gb)(-1+d gb)(1+d gb)),"\\
\verb"S1->((-1+g)(1+g)gb^2(-1+d gb)(1+d gb))/(g^2(-1+d g)(1+d g)(-1+gb)(1+gb)),"\\
\verb"S2->((-1+g)(1+g)(-1+gb)(1+gb))/(g^2(-1+d g)(1+d g)gb^2(-1+d gb)(1+d gb)),"\\
\verb"P-> ((-1+d g^2)^2(-1+d gb^2)^2)/((-1+d)^4g^2gb^2),"\\
\verb"T2->d g gb, T1->-(g/gb)};"\\
\\
\verb"Ym[a_]=-1+(S2 T2^(1+a)(-1+T2^2)-S1^2 S2 T1^(4+2a)T2^(1+a)(-1+T2^2)+"\\
\verb"S1 T1^(1+a)(-1+T1^2)(-1+S2^2 T2^(4+2a)))^2/"\\
\verb"((-S2 T2^a (-1+T2^2)+S1^2 S2 T1^(2+2 a)T2^a (-1+T2^2)-"\\
\verb"S1 T1^a (-1+T1^2)(-1+S2^2 T2^(2+2 a)))(-S2 T2^(2+a)(-1+T2^2)+"\\
\verb"S1^2 S2 T1^(6+2 a) T2^(2+a)(-1+T2^2)-"\\
\verb"S1 T1^(2+a)(-1+T1^2)(-1+S2^2 T2^(6+2 a))))/.sb;"\\
\\
\verb"Yp[a_]=((S1 T1^(4+a) T2)/(-1+T1^2)+(T1^-a T2)/(S1-S1 T1^2)-"\\
\verb"(T1(T2^-a-S2^2 T2^(4+a)))/(S2 - S2 T2^2))^2/(-((S1 T1^(4+a)T2)/"\\
\verb"(-1+T1^2)+(T1^-a T2)/(S1 - S1 T1^2)-(T1(T2^-a-S2^2 T2^(4+a)))/"\\
\verb"(S2-S2 T2^2))^2+(T1^(-2 a)T2^(-2 a)(T2^2-S2^2 T2^(4+2 a)-"\\
\verb"2 S1 S2 T1^(2+a)T2^(2+a)(-1+T2^2)+2 S1 S2 T1^(4+a)T2^(2+a)(-1+T2^2)+"\\
\verb"S1^2 T1^(6+2 a)T2^2(-1+S2^2 T2^(2+2 a))+2 T1 T2 (-1+S2^2 T2^(4+2 a))-"\\
\verb"2 S1^2 T1^(5+2 a)T2(-1+S2^2 T2^(4+2 a))+T1^2 (1-S2^2 T2^(6+2 a))+"\\
\verb"S1^2 T1^(4+2 a)(-1+S2^2 T2^(6+2 a)))^2)/(S1^2 S2^2 (-1+T1^2)^2"\\
\verb"(-1+T2^2)^2(T2+T1^2 T2-T1(1+T2^2))^2))/.sb;"\\
\\
\verb"Yb[s_]=(s-A)^2-1/.sb;"\\
\\
\verb"Y11=(-1+(S1 S2(-1+T1^2)(-1+T2^2)(T2^2-S2^2 T2^6-2 S1 S2 T1^3 T2^3 (-1+T2^2)+"\\
\verb"2 S1 S2 T1^5 T2^3(-1+T2^2)+S1^2 T1^8 T2^2(-1+S2^2 T2^4)+"\\
\verb"2 T1 T2(-1+S2^2 T2^6)-2 S1^2 T1^7 T2(-1+S2^2 T2^6)+T1^2 (1-S2^2 T2^8)+"\\
\verb"S1^2 T1^6 (-1+S2^2 T2^8))^2)/((S1^2 S2 T1^4 T2 (-1+T2^2)+S2 (T2-T2^3)-"\\
\verb"S1 T1 (-1+T1^2)(-1+S2^2 T2^4))^2 (T2^2-S2^2 T2^8-2 S1 S2 T1^4 T2^4 (-1+T2^2)+"\\
\verb"2 S1 S2 T1^6 T2^4 (-1+T2^2)+S1^2 T1^10 T2^2 (-1+S2^2 T2^6)+2 T1 T2 (-1+S2^2 T2^8)-"\\
\verb"2 S1^2 T1^9 T2 (-1+S2^2 T2^8)+T1^2 (1-S2^2 T2^10)+S1^2 T1^8 (-1+S2^2 T2^10))))/.sb;"\\
\\
\verb"Y22=(P/Y11)/.sb;"\\


    \end{document}